\documentclass[useAMS,usenatbib,usegraphicx]{mn2e}
\usepackage{bm}
\usepackage{times}

\title[Evolution of grain sizes in galaxies]
{Two-size approximation: a simple way of treating
the evolution of grain size distribution in galaxies}
\author[Hirashita]{Hiroyuki Hirashita\thanks{E-mail:
    hirashita@asiaa.sinica.edu.tw}\\
Institute of Astronomy and Astrophysics, Academia Sinica,
PO Box 23-141, Taipei 10617, Taiwan\\
}
\date{2014 December 11}

\pagerange{\pageref{firstpage}--\pageref{lastpage}} \pubyear{2014}

\begin{document}
\label{firstpage}
\maketitle

\begin{abstract}
Full calculations of the evolution of grain size distribution
in galaxies are in general computationally heavy. In this paper,
we propose a simple model of dust enrichment in a galaxy with
a simplified treatment of grain size distribution by imposing
a `two-size approximation'; that is, all the grain population is
represented by small (grain radius $a<0.03~\micron$) and
large ($a>0.03~\micron$)
grains. We include in the model dust supply from
stellar ejecta, destruction in supernova shocks, dust growth
by accretion, grain growth by coagulation and grain disruption by
shattering, considering how these processes work on the small
and large grains. We show that this simple framework reproduces the
main features found in full calculations of grain size distributions
as follows.
The dust enrichment starts with the supply of large grains from
stars. At a metallicity level referred to as the critical metallicity
of accretion, the abundance of the small grains formed by shattering
becomes large enough to rapidly increase the grain abundance by
accretion. Associated with this epoch, the mass ratio of the small grains
to the large grains reaches the maximum. After that, this ratio
converges to the value determined by the balance between
shattering and
coagulation, and the dust-to-metal ratio is determined by the
balance between accretion and shock destruction. With a Monte
Carlo simulation, we demonstrate
that the simplicity
of our model has an advantage in predicting statistical properties.
We also show some applications for predicting observational dust
properties such as extinction curves.
\end{abstract}


\section{Introduction}

Dust enrichment has a large impact on the evolution of
galaxies in various aspects. Dust grains induce the formation
of molecular gas through dust surface reaction of some
molecular species, especially H$_2$
\citep[e.g.][]{gould63,hirashita02,cazaux04}.
Dust also affects the typical stellar mass or the initial
mass function (IMF) \citep{schneider06} as dust cooling
induces fragmentation of molecular clouds
\citep{omukai00,omukai05}. Moreover, dust grains absorb optical
and ultraviolet (UV) stellar light and reprocess it into infrared
bands,  dramatically affecting the observed galaxy spectral
energy distributions (SEDs) \citep[e.g.][]{takeuchi05}.
The efficiencies of the above processes are mostly proportional
to the total dust amount (or abundance) in the system.

Besides the total dust amount, dust properties constitute
another important aspect in dust evolution.
Among them, the grain size distribution is
of particular importance since the optical properties
such as extinction curves and grain-surface
chemical reaction rates are directly modified if the
grain size distribution changes. In other words,
even if we precisely modeled the evolution of total
dust amount in a system, lack of knowledge of the
grain size distribution would cause large uncertainties
in optical and chemical characteristics of the system.
Moreover, as shown below, the grain size distribution
also affects the evolution of the total dust amount
\citep[see also][]{kuo12}. Therefore,
it is crucial to model the total dust amount and the dust
properties (especially the grain size distribution) at the
same time.

Recently \citet[][hereafter A13]{asano13a} have constructed
a full framework of treating the evolution of grain size
distribution over the
entire galaxy evolution. Their calculations suggest the
following evolutionary features of grain size distribution.
At the early stage of galaxy evolution, when the system is
metal-poor, the dust is predominantly supplied from
core-collapse supernovae (simply referred to as SNe)
and asymptotic giant branch (AGB) stars.
If the system reaches a metallicity level referred to as
the critical metallicity, shattering produces a large amount of
small grains, which eventually grow by accreting gas-phase
metals in the dense interstellar medium (ISM). This
accretion becomes the dominant mechanism of grain mass
increase at metallicities beyond the critical metallicity.
After that, coagulation
in dense clouds pushes the small grains towards larger sizes.
The evolution of extinction curves
is also investigated by \citet{asano14} using the evolution models
of grain size distribution in A13.

The importance of including dust in cosmological galaxy
evolution models has been shown by many authors. Among them,
\citet*{dayal10} included
dust production and destruction in a cosmological simulation,
predicting the statistical properties of dust emission
luminosities in a large sample of high-redshift galaxies.
More elaborate dust evolution models that include
two major dust formation mechanisms (dust formation in stellar
ejecta and dust growth by accretion in the ISM) are also
incorporated in semi-analytic
{\citep{valiante11,debennassuti14}} or
$N$-body \citep{bekki13,yozin14} frameworks.

However, the above cosmological galaxy evolution models only take
the total dust content into account. It still remains challenging
to include grain size distributions in these
frameworks, because the freedom of grain
size adds another dimension to the calculations. In fact,
\citet{yajima14} show that the extinction properties
of Lyman-break galaxies are influenced by the assumed dust
size. At the same time, as shown by A13 (see above), the evolution of
the total dust amount is coupled with that of the grain size
distribution.
Therefore, it is desirable to develop a method of treating
the information on grain size distribution in a computationally
light manner.

The first purpose of this paper is to develop a simple and light
method of treating the evolution of grain size distribution in a
galaxy. Such a `light' method is also
useful to analyze
the dependence on various parameters that govern the grain
size distribution, since we can easily run numerous cases
with different parameter values. Thus, the second purpose
of this paper is detailed analysis of the response of dust
enrichment to various dust enrichment and processing
mechanisms. We also address future prospect of including
the light dust evolution model developed in this paper
into such complex galaxy evolution models as mentioned above
(i.e.\ semi-analytic models, $N$-body simulations, etc.).

In this paper, we first review the dust enrichment and
processing mechanisms in Section \ref{subsec:review},
where we note that most of the dust formation and processing
mechanisms work distinctively on small
($\la 0.03~\micron$) grains and large ($\ga 0.03~\micron$)
grains. This motivates us to apply a
`two-size approximation', in which we represent the grain
size distribution by the ratio of the total mass of
the small grains to that of the large grains. This ratio is
referred to as the `small-to-large grain abundance ratio'.
As demonstrated in this paper,
this `two-size approximation' nicely catches the features in
the full calculations of grain size evolution. We also
propose to use this two-size approximation in complex galaxy
evolution models.

The paper is organized as follows: we review the processes to be
taken into account, and formulate the model by using the
two-size approximation in Section~\ref{sec:model}. We show the
results in Section~\ref{sec:result}. We analyze and discuss
the results in Section \ref{sec:discussion}
and consider possibilities of including our models in complex
galaxy evolution models in Section \ref{sec:prospect}.
Finally we conclude in Section \ref{sec:conclusion}.

\section{Model}\label{sec:model}

\subsection{Review of the processes}\label{subsec:review}

In considering the evolution of dust in the ISM of a galaxy,
we need to consider not only the dust supply from stars
but also dust processing in the ISM. Following A13,
we include the
following processes considered to dominate the evolution
of dust content and grain size distribution:
(i) dust supply from stellar ejecta, (ii) dust destruction in
the ISM by supernova (SN) shocks, (iii) grain growth by the accretion of
gas-phase metals, (iv) grain growth by coagulation (sticking of
grains), and (v) shattering
(grain disruption/fragmentation). These processes act differently
on large and small grains. Here we roughly divide the grains into
large and small grains at $a\sim 0.03~\micron$, where $a$ is the
grain radius. We hereafter refer to grains with $a>0.03~\micron$
and $a<0.03~\micron$
as the large grains and the small grains, respectively.
We summarize these processes in Table~\ref{tab:summary}.
We also show, based on A13, how each process acts on different grain
sizes. We explain each process in what follows.

\begin{table}
\centering
\begin{minipage}{80mm}
\caption{Processes considered in this paper.}
\label{tab:summary}
\begin{center}
\begin{tabular}{@{}lccc} \hline
Process$^a$ & Small Grains$^b$ & Large Grains$^b$ & Total
\\ \hline
Stellar ejecta & $\triangle$ & $\nearrow$ & $\nearrow$ \\
Shock destruction & $\searrow$ & $\searrow$ & $\searrow$ \\
Accretion & $\nearrow$ & $\triangle$ & $\nearrow$ \\
Coagulation & $\searrow$ & $\nearrow$ & $\longrightarrow$ \\
Shattering & $\nearrow$ & $\searrow$ & $\longrightarrow$ \\ \hline
\end{tabular}
\end{center}

Note:
$\nearrow$, $\searrow$, and $\longrightarrow$ indicate that
the process increases, decreases, and keeps constant
the dust mass, respectively. $\triangle$ means that the
process has little influence on the dust mass.\\
$^{a}$Each process is described in Section \ref{subsec:review}.\\
$^{b}$The large and small grains are divided at $a\sim 0.03~\micron$.
\end{minipage}
\end{table}

The dominant stellar sources of dust are SNe and
AGB stars \citep[][for a review]{gall11}, and both types
of sources are predicted to supply
large grains. The dust supplied from SNe tends to be biased
to large sizes because reverse shock destruction is more
effective for small grains than large ones
(\citealt{nozawa07}; see also \citealt{bianchi07}).
The typical size of grains produced
by AGB stars is also suggested to be large
($a\ga 0.1~\mu$m) from observations of
SEDs \citep{groenewegen97,gauger99},
although \citet{hofmann01} showed that the grains are not
single-sized. Theoretical studies have also shown that
large ($\ga 0.1~\micron$)
dust grains form in the winds of AGB stars
\citep{winters97,yasuda12,ventura12}.
There are also pieces of evidence obtained by
meteoritic samples that dust species such as SiC
thought to originate from AGB stars from
isotopic compositions
\citep{hoppe94,daulton03} have large grain sizes
($a\ga 0.1~\mu$m), supporting the formation of large
grains in AGB stars \citep{amari94,hoppe00}.

Dust grains, after being injected into the ISM, are
destroyed by sputtering if they are swept by SN shocks
\citep{dwek80,mckee89}. Because thermal sputtering is
a surface process (a process whose rate is proportional to the
total surface area), the dust destruction time-scale by this
process is proportional to the grain radius \citep{draine79}.
Nonthermal sputtering has little dependence
on the grain size \citep{jones96}. In essence, for sputtering,
it is difficult to separate large and small
grains, since `large' grains being destroyed enter
the small regime. Therefore, to simplify the treatment and
to minimize the number of parameters, we apply
the same time-scale of destruction both for the large and
the small grains unless otherwise stated. We also examine
the case in which the small grains are more efficiently
destroyed.

Another important dust formation process is grain growth
by the accretion of gas-phase metals in the dense ISM
{\citep{dwek98,hirashita99,zhukovska08,draine09,inoue11,pipino11,asano13b}}.
This process is simply referred to as accretion in this paper.
Including accretion into dust evolution models is motivated not
only to explain the dust abundance in nearby galaxies
{\citep{hirashita99,inoue11,zhukovska13,debennassuti14,schneider14}},
but also to explain the observations of huge amounts of dust
($\sim 10^8$ M$_\odot$) in
high-$z$ quasars and starbursts
\citep{michalowski10a,mattsson11,valiante11,kuo12,rowlands14,nozawa14}.
Grain growth by accretion occurs efficiently only after
the ISM is significantly enriched with dust and metals, since
the grain growth rate is proportional to the collision rate
between these two components.
{Since the small grains have much larger surface-to-volume
ratios than the large grains,
accretion has a predominant influence on the small grains,
creating a bump in the grain size distribution at $a<0.03~\micron$
\citep{hirashita11,hirashita12}. The grain size distribution
is hardly changed by accretion at $a>0.03~\micron$.}

Coagulation occurs in the dense ISM and pushes the grain size
distribution toward larger grain sizes
\citep[e.g.][]{ormel09,hirashita14}. In particular, coagulation is
a unique mechanism that converts small grains to
large grains, since another grain growth mechanism, accretion,
cannot increase the grain radius drastically
{as mentioned above}. Since coagulation converts small
grains to large grains, it can be regarded as a destruction
mechanism for the small grains and as a formation mechanism for
the large grains.

Shattering occurs predominantly in the diffuse ISM
\citep{yan04,hirashita09} and in SN shocks \citep{jones96},
and creates a large number of small grains from large grains.
Shattering is
{the most efficient mechanism of producing small grains since
the fragmentation associated with shattering produce a large
number of small grains. Shattering}
can be regarded as a destruction mechanism for
the large grains and as a formation mechanism for the small grains.

\subsection{Evolution of small and large grains}\label{subsec:evol}

We include the above processes in a simple evolution model of
dust enrichment in a galaxy.
The purpose of our model is to construct a light and simple
evolution model of dust amount and grain size, which can be
easily included in computationally heavy frameworks of
galaxy evolution such
as $N$-body and/or smoothed particle hydrodynamics (SPH) simulations
and semi-analytic models.
Therefore, we keep our model in this paper as simple as possible
by adopting a one-zone closed-box model described in
\citet{hirashita11}, although extension of our framework to
more elaborate treatments is straightforward
\citep{dwek98,lisenfeld98}.

The model treats the evolutions of the total gas, metal and dust masses
($M_\mathrm{g}$, $M_Z$ and $M_\mathrm{d}$, respectively) in the galaxy.
In this model, the metals include not only gas phase elements but also
dust. $M_\mathrm{g}$ includes $M_Z$ and $M_\mathrm{d}$,
but $M_\mathrm{g}\gg M_Z$ and $M_\mathrm{d}$ in any case. We newly
divide the dust component into the small and large grains
(Section~\ref{subsec:review}), whose total masses are denoted as
$M_\mathrm{d,s}$ and $M_\mathrm{d,l}$,
respectively. Considering the processes described in
Section \ref{subsec:review} and listed in Table \ref{tab:summary},
the equations are written as
\begin{eqnarray}
\frac{\mathrm{d}M_\mathrm{g}}{\mathrm{d}t} & = &
-\psi +E,\label{eq:dMgdt}\\
\frac{\mathrm{d}M_{Z}}{\mathrm{d}t} & = &
-Z\psi+E_{Z}\\
\frac{\mathrm{d}M_\mathrm{d,s}}{\mathrm{d}t} & = &
-\mathcal{D_\mathrm{s}}\psi -\frac{M_\mathrm{d,s}}{\alpha\tau_\mathrm{SN}}+
\frac{M_\mathrm{d,l}}{\tau_\mathrm{sh}}
-\frac{M_\mathrm{d,s}}{\tau_\mathrm{co}}
+\frac{M_\mathrm{d,s}}{\tau_\mathrm{acc}},\label{eq:dMdsdt}\\
\frac{\mathrm{d}M_\mathrm{d,l}}{\mathrm{d}t} & = &
-\mathcal{D_\mathrm{l}}\psi +f_\mathrm{in}E_Z-
\frac{M_\mathrm{d,l}}{\tau_\mathrm{SN}}-
\frac{M_\mathrm{d,l}}{\tau_\mathrm{sh}}+
\frac{M_\mathrm{d,s}}{\tau_\mathrm{co}},\label{eq:dMdldt}
\end{eqnarray}
where $\psi$ is the star formation rate, $E$ and $E_Z$ are
the rates of the total injection of mass (gas + dust) and
metal mass from stars, respectively, and $f_\mathrm{in}$
is the dust condensation efficiency of the metals in
the stellar ejecta. The time-scales of various processes
are also introduced: $\tau_\mathrm{SN}$ is the
time-scale of dust destruction by SN shocks for the large grains,
and $\alpha\tau_\mathrm{SN}$ is that for the small grains
(i.e.\ $\alpha\leq 1$ is introduced to consider a possible
short destruction time for small grains; Section \ref{subsec:review};
\citealt{nozawa06}); $\tau_\mathrm{sh}$,
$\tau_\mathrm{co}$, and $\tau_\mathrm{acc}$ are the
time-scales of shattering, coagulation, and accretion,
respectively. We also define the metallicity,
$Z\equiv M_Z/M_\mathrm{g}$, the small-grain dust-to-gas ratio,
$\mathcal{D}_\mathrm{s}\equiv M_\mathrm{d,s}/M_\mathrm{g}$,
the large-grain dust-to-gas ratio,
$\mathcal{D}_\mathrm{l}\equiv M_\mathrm{d,l}/M_\mathrm{g}$,
and the dust-to-gas ratio,
$\mathcal{D}\equiv\mathcal{D}_\mathrm{d,s}+\mathcal{D}_\mathrm{d,l}$.

In order to allow a simple analytical treatment, we adopt the
instantaneous recycling approximation; that is, a star with $m>m_t$
($m$ is the zero-age stellar mass, and $m_t$ is
the turn-off mass at fixed age $t$) dies
instantaneously after its birth, leaving a remnant of
mass $w_m$ \citep{tinsley80}. Once the initial mass function (IMF) is
fixed, the returned fraction of the mass
from formed stars, $\mathcal{R}$, and the
mass fraction of metals that is newly produced
and ejected by stars, $\mathcal{Y}_{Z}$, are evaluated.
Using these quantities, we write
\begin{eqnarray}
E & = & \mathcal{R}\psi ,\\
E_{Z} & = & (\mathcal{R}Z+
\mathcal{Y}_{Z})\psi .
\end{eqnarray}

For the time-scale of large grain destruction by SN shocks, we
define a parameter,
$\beta_\mathrm{SN}\equiv\tau_\mathrm{SF}/\tau_\mathrm{SN}$,
where $\tau_\mathrm{SF}\equiv M_\mathrm{g}/\psi$ is
the star formation time-scale (treated as a constant for
simplicity in this paper) and
$\tau_\mathrm{SN}=M_\mathrm{g}/(\epsilon_\mathrm{sw,l}
M_\mathrm{sw}\gamma )$ ($\epsilon_\mathrm{sw,l}$
and $M_\mathrm{sw}$ are the dust destruction efficiency
and the gas mass swept by a single high-velocity SN
blast, respectively, and $\gamma$ is the SN rate).
With the above definitions,
$\beta_\mathrm{SN}=\epsilon_\mathrm{sw,l}M_\mathrm{sw}\gamma /\psi$,
which is treated as a constant with the instantaneous
recycling approximation \citep{hirashita11}.

The time-scales of shattering and coagulation depend on
the dust-to-gas ratio, since
these processes become efficient as the system is enriched with
dust. In our models, we consider shattering of the large
grains into the small grains and coagulation of the small grains
into the large grains, assuming the following dependences
for these two time-scales:
\begin{eqnarray}
\tau_\mathrm{sh} & = & \tau_\mathrm{sh,0}\left(
\frac{\mathcal{D}_\mathrm{l}}{\mathcal{D}_\mathrm{MW,l}}
\right)^{-1},\label{eq:tau_shat}\\
\tau_\mathrm{co} & = & \tau_\mathrm{co,0}\left(
\frac{\mathcal{D}_\mathrm{s}}{\mathcal{D}_\mathrm{MW,s}}
\right)^{-1},\label{eq:tau_coag}
\end{eqnarray}
where we normalize the time-scales to the values
($\tau_\mathrm{sh,0}$ and $\tau_\mathrm{co,0}$) at the Milky Way
dust-to-gas ratios 
and the total Milky Way dust-to-gas ratio, 0.01, is divided
into the large-grain dust-to-gas ratio
($\mathcal{D}_\mathrm{MW,l}=0.0070$) and
the small-grain dust-to-gas ratio
($\mathcal{D}_\mathrm{MW,s}=0.0030$) assuming
the power-law grain size distribution ($\propto a^{-3.5}$) in
the grain size ranges 0.001--0.25 $\micron$, which is
applicable for the Milky Way dust \citep*[][hereafter MRN]{mathis77}.
For shattering and coagulation, we define $\beta_\mathrm{sh}$
and $\beta_\mathrm{co}$ in similar ways to $\beta_\mathrm{SN}$
above:
$\beta_\mathrm{sh}\equiv\tau_\mathrm{SF}/\tau_\mathrm{sh}$
and $\beta_\mathrm{co}\equiv\tau_\mathrm{SF}/\tau_\mathrm{co}$.
Note that $\beta_\mathrm{sh}$ and $\beta_\mathrm{co}$ varies
in proportion to $\mathcal{D}_\mathrm{l}$ and $\mathcal{D}_\mathrm{s}$,
respectively, while $\beta_\mathrm{SN}$
is treated as a constant.

For accretion, we adopt the formulation developed by
\citet{hirashita11}, who also considered the dependence on
grain size distribution.
The increasing rate of dust mass by accretion is expressed as
\begin{eqnarray}
\frac{M_\mathrm{d,s}}{\tau_\mathrm{acc}}
=\frac{\mathcal{B}X_\mathrm{cl}M_\mathrm{d,s}}{\tau_\mathrm{cl}},\label{eq:dmdt_acc}
\end{eqnarray}
where we introduce the accretion time-scale, $\tau_\mathrm{acc}$,
$X_\mathrm{cl}$ is the cold cloud fraction to the total
gas mass, $\tau_\mathrm{cl}$ is the lifetime of the cold clouds, and
$\mathcal{B}$ is the increment of dust mass in the cold
clouds\footnote{\citet{hirashita11} used the notation $\beta$ for
$\mathcal{B}$.
In order to avoid confusion with $\beta_\mathrm{acc}$ introduced
later, we use $\mathcal{B}$ in this paper.}, which can
be estimated as
\begin{eqnarray}
\mathcal{B}\simeq\left[
\frac{\langle a^3\rangle_0}{3y\langle a^2\rangle_0+3y^2\langle a\rangle_0
+y^3}
+\frac{\mathcal{D}_\mathrm{s}}{Z-\mathcal{D}}\right]^{-1},\label{eq:beta}
\end{eqnarray}
where
$y\equiv a_0\xi\tau_\mathrm{cl}/\tau$ ($a_0$ is just used
for normalization,
{$\xi$ is the fraction of metals in gas phase,}
and $\tau$ is the accretion time-scale for a grain with
radius $a_0$), and
$\langle a^\ell\rangle_0$ is the $\ell$th moment of grain radius.
The term $\mathcal{D}_\mathrm{s}/(Z-\mathcal{D})$ is based on the
fact that $\mathcal{B}$ cannot be larger than the case in which all the
gas-phase metals are used up
(i.e.\ $\mathcal{B}<(Z-\mathcal{D})/\mathcal{D}_\mathrm{s}$).
{Here we implicitly assume that the accretion rate in the cold clouds
is not affected by other grain processing mechanisms. In fact,
coagulation also occurs there, but \citet{hirashita12} showed
that coagulation does not alter the dust mass growth rate
by accretion (or $\mathcal{B}$).}
We adopt the following expression for $\tau$
(see eq.~23 of \citealt{hirashita11}, applicable for silicate, but carbonaceous dust
has a similar time-scale):
\begin{eqnarray}
\tau=6.3\times 10^7\left(\frac{Z}{\mathrm{Z}_{\sun}}\right)^{-1}
a_{0.1}n_3^{-1}T_{50}^{-1/2}S_{0.3}^{-1}~\mathrm{yr},\label{eq:tau}
\end{eqnarray}
where $a_{0.1}\equiv a_0/(0.1~\micron )$,
$n_3\equiv n_\mathrm{H}/(10^3~\mathrm{cm}^{-3})$
($n_\mathrm{H}$ is the number density of hydrogen nuclei in
the cold clouds),
$T_\mathrm{50}\equiv T_\mathrm{gas}/(50~\mathrm{K})$ ($T_\mathrm{gas}$
is the gas temperature), and $S_{0.3}\equiv S/0.3$ ($S$ is the
sticking probability of the dust-composing material onto the preexisting
grains).
We use the same values as in \citet{hirashita11}; i.e.\ $a_0 = 0.1~\micron$,
$n_\mathrm{H} = 10^3$ cm$^{-3}$, $T_\mathrm{gas} = 50$ K, and $S = 0.3$.
Assuming that the cold clouds hosting accretion and star formation
are the same, we can express the star formation rate as
\begin{eqnarray}
\frac{M_\mathrm{g}}{\tau_\mathrm{SF}}=
\frac{\epsilon X_\mathrm{cl}M_\mathrm{g}}{\tau_\mathrm{cl}},
\label{eq:SFR}
\end{eqnarray}
where $\epsilon$ is the star formation efficiency of the
cold clouds. We define
$\beta_\mathrm{acc}\equiv\tau_\mathrm{SF}/\tau_\mathrm{acc}$,
which by using equations (\ref{eq:dmdt_acc}) and (\ref{eq:SFR})
can be evaluated as
\begin{eqnarray}
\beta_\mathrm{acc}=\frac{\mathcal{B}}{\epsilon}.
\end{eqnarray}

Equations (\ref{eq:dMgdt})--(\ref{eq:dMdldt}) are converted to the time
evolution of the metallicity $Z=M_Z/M_\mathrm{g}$, the small-grain
dust-to-gas
ratio $\mathcal{D}_\mathrm{s}=M_\mathrm{d,s}/M_\mathrm{g}$, and
the large-grain dust-to-gas ratio
$\mathcal{D}_\mathrm{l}=M_\mathrm{d,l}/M_\mathrm{g}$ as
\begin{eqnarray}
\frac{M_\mathrm{g}}{\psi}\frac{\mathrm{d}Z}{\mathrm{d}t} & = & \mathcal{Y}_Z,
\label{eq:dZdt}\\
\frac{M_\mathrm{g}}{\psi}\frac{\mathrm{d}\mathcal{D}_\mathrm{l}}{\mathrm{d}t}
& = & f_\mathrm{in}(\mathcal{R}Z+\mathcal{Y}_Z)+
\beta_\mathrm{co}\mathcal{D}_\mathrm{s}\nonumber\\
& & -
(\beta_\mathrm{SN}+\beta_\mathrm{sh}+\mathcal{R})
\mathcal{D}_\mathrm{l},\\
\frac{M_\mathrm{g}}{\psi}\frac{\mathrm{d}\mathcal{D}_\mathrm{s}}{\mathrm{d}t}
& = & \beta_\mathrm{sh}\mathcal{D}_\mathrm{l}-
\left(\frac{\beta_\mathrm{SN}}{\alpha}+\beta_\mathrm{co}+\mathcal{R}
-\beta_\mathrm{acc}\right)
\mathcal{D}_\mathrm{s}.\label{eq:dDsdt}
\end{eqnarray}
By rearranging equations (\ref{eq:dZdt})--(\ref{eq:dDsdt}), we obtain
\begin{eqnarray}
\mathcal{Y}_Z\frac{\mathrm{d}\mathcal{D}_\mathrm{l}}{\mathrm{d}Z} & = &
f_\mathrm{in}(\mathcal{R}Z+\mathcal{Y}_Z)+
\beta_\mathrm{co}\mathcal{D}_\mathrm{s}\nonumber\\
& & -
(\beta_\mathrm{SN}+\beta_\mathrm{sh}+\mathcal{R})
\mathcal{D}_\mathrm{l},\label{eq:dDldt2}\\
\mathcal{Y}_Z\frac{\mathrm{d}\mathcal{D}_\mathrm{s}}{\mathrm{d}Z} & = &
\beta_\mathrm{sh}\mathcal{D}_\mathrm{l}-
\left(\frac{\beta_\mathrm{SN}}{\alpha} +\beta_\mathrm{co}+\mathcal{R}-
\beta_\mathrm{acc}\right)\mathcal{D}_\mathrm{s}.\label{eq:dDsdt2}
\end{eqnarray}
These two equations are solved to obtain the
$\mathcal{D}_\mathrm{l}$--$Z$ and $\mathcal{D}_\mathrm{s}$--$Z$
relations.
If we add Equations (\ref{eq:dDldt2}) and (\ref{eq:dDsdt2}),
we obtain the evolution of total dust-to-gas ratio,
$\mathcal{D}=\mathcal{D}_\mathrm{l}+\mathcal{D}_\mathrm{s}$, as
\begin{eqnarray}
\mathcal{Y}_Z\frac{\mathrm{d}\mathcal{D}}{\mathrm{d}Z}=
f_\mathrm{in}(\mathcal{R}Z+\mathcal{Y}_Z)-
(\beta_\mathrm{SN}' +\mathcal{R})\mathcal{D}
+\beta_\mathrm{acc}\mathcal{D}_\mathrm{s},\label{eq:dg_metal}
\end{eqnarray}
where $\beta_\mathrm{SN}'\equiv\beta_\mathrm{SN}
(\mathcal{D}_\mathrm{l}+\mathcal{D}_\mathrm{s}/\alpha)/\mathcal{D}$
(if $\alpha =1$, $\beta_\mathrm{SN}'=\beta_\mathrm{SN}$).
Equation (\ref{eq:dg_metal}) is a similar equation to eq.\ (48)
of \citet{hirashita11} except that the mass increase by
accretion is only considered for the small grains in the current paper.
The processes that conserve the total dust mass, that is,
coagulation and shattering, do not explicitly appear in the evolution
of $\mathcal{D}$, but implicitly affect the evolution of
$\mathcal{D}_\mathrm{s}$ in the last term.

\subsection{Choice of parameter values}
\label{subsec:parameter}

As mentioned {in the beginning of Section \ref{subsec:review}},
we set the boundary of the small
and large grains at $a=0.03~\micron$.
{Indeed, this value can be justified by A13's full calculation of
grain size distribution: in their fig.\ 6 (left panel),
the processes dominating the abundance of small grains
(accretion, coagulation, and shattering) create a bump in the
size distribution at small grain sizes. At large grain sizes,
there is another bump cerated by the dust production by stars.
The boundary of these two bumps are around $a\simeq 0.03~\micron$,
representing the different processes between small and large
grains for the evolution of grain size distribution. Therefore,
we adopt $a=0.03~\micron$ for the boundary.}
For accretion, we need to assume a
grain size distribution to evaluate the moments of
grain radius in equation (\ref{eq:beta}). We assume
the MRN grain size distribution ($\propto a^{-3.5}$)
between $a=0.001$ and 0.03 $\micron$ for the small grains, obtaining
$\langle a\rangle_0=1.66\times 10^{-3}~\micron$,
$\langle a^2\rangle_0{}^{1/2}=2.02\times 10^{-3}~\micron$, and
$\langle a^3\rangle_0{}^{1/3}=2.82\times 10^{-3}~\micron$.
For accretion, the `volume-to-surface ratio',
$\langle a^3\rangle_0/\langle a^2\rangle_0=5.48\times 10^{-3}~\micron$
is the most important quantity, and this value is near
to the bump created by accretion (A13), which confirms the
validity of the moments that we adopted.

We adopt $\mathcal{R}=0.25$ and
$\mathcal{Y}_Z=0.013$ \citep*{hirashita11,kuo13}
under a Salpeter IMF with a stellar mass range
of 0.1--100 M$_{\sun}$ and the instantaneous recycling
for $t=5$ Gyr.
Although we can change the values of these parameters
by adopting a different IMF, etc., we fix them, since they
have only a minor influence on the relation between
dust-to-gas ratio and metallicity compared with
the values of $f_\mathrm{in}$ and parameters related to
accretion.

For the fiducial case, we choose $\tau_\mathrm{SF}=5$ Gyr,
which is roughly appropriate
for nearby spiral galaxies (we also examine $\tau_\mathrm{SF}=0.5$
and 50 Gyr). Since we treat $\tau_\mathrm{SF}$ as a constant,
we cannot include the effects of episodic star formation,
which may enhance the scatter of the relation between dust-to-gas
ratio and metallicity \citep{zhukovska14}.
For $f_\mathrm{in}$, we adopt
$f_\mathrm{in}=0.1$ following \citet{inoue11}, but we
also apply their lower efficiency case,  $f_\mathrm{in}=0.01$.
We adopt $\beta_\mathrm{SN}=9.65$ following
\citet{hirashita11}. We also vary $\beta_\mathrm{SN}$ by a
factor of 2. Such a factor 2 variation is
expected by considering inhomogeneity or multi-phase
structures in the ISM \citep{mckee89}.
\citet{jones11} pointed out that the estimate of
$\tau_\mathrm{SN}$ is uncertain; however,
as shown later, our choice of
$\beta_\mathrm{SN}$ nicely explains the dust-to-metal
ratio of the Milky Way (Section~\ref{subsec:highZ}).
For the destruction of small grains, we
conservatively choose $\alpha =1$, but later we will also
discuss the cases with $\alpha <1$. As noted above,
$\beta_\mathrm{SN}=\tau_\mathrm{SF}/\tau_\mathrm{SN}$ can
be treated as a constant under the instantaneous recycling
approximation.

Accretion is regulated by
$\beta_\mathrm{acc}=\mathcal{B}/\epsilon$ in
equation (\ref{eq:beta}), which is parameterized by
the cloud
lifetime $\tau_\mathrm{cl}$ and star formation
efficiency $\epsilon$.
{Because of the degeneracy,
we only explore variations of the parameter $\tau_\mathrm{cl}$
with fixed values of the other parameters in $\mathcal{B}$}:
as we observe in
equation (\ref{eq:beta}),
a large/small value of $\tau_\mathrm{cl}$ indicates
efficient/inefficient accretion;
this is because the lifetime of a cloud hosting
accretion regulates the duration of accretion there.
We adopt $\tau_\mathrm{cl}=10^7$ yr for the fiducial value
\citep{hirashita11}
with an order of magnitude variation also investigated.
For the star formation efficiency in the cold clouds,
we adopt $\epsilon =0.1$.

\citet*{seok14} derived shattering and coagulation time-scales
in their application of PAH formation and destruction, and we apply
approximately the same values for the time-scales for the fiducial
case; that is, we adopt
$\tau_\mathrm{sh,0}=10^8$ yr and
$\tau_\mathrm{co,0}= 10^7$ yr {for the fiducial case}.
Since we expect large variations for both parameters
depending on the ISM density and the fractions of various ISM
phases (A13), we consider an order of magnitude variations for
them.

The above choice of the parameters are summarized in
Table~\ref{tab:param}.
For the normalization of metallicity,
we adopt the same solar metallicity Z$_{\sun}=0.02$ as
adopted in A13 \citep[originally from][]{anders89}.

\begin{table}
\centering
\begin{minipage}{80mm}
\caption{Parameter ranges surveyed.}
\label{tab:param}
\begin{center}
\begin{tabular}{@{}llccc} \hline
Parameter & Process & Minimum & Maximum & Fiducial
\\ \hline
$f_\mathrm{in}$ & stellar ejecta & 0.01 & 0.1 & 0.1\\
$\beta_\mathrm{SN}$ & SN destruction & 4.8 & 19 & 9.65\\
$\tau_\mathrm{cl}$ & accretion & $10^6$ yr & $10^8$ yr & $10^7$ yr\\
$\tau_\mathrm{sh,0}$ & shattering & $10^7$ yr & $10^9$ yr & $10^8$ yr\\
$\tau_\mathrm{co,0}$ & coagulation & $10^6$ yr & $10^8$ yr & $10^7$ yr\\
$\tau_\mathrm{SF}$ & star formation & 0.5 Gyr & 50 Gyr & 5 Gyr\\ \hline
\end{tabular}
\end{center}
\end{minipage}
\end{table}

\section{Results}\label{sec:result}

\subsection{Relation between dust-to-gas ratio and metallicity}
\label{subsec:dg_metal}

\begin{figure}
\includegraphics[width=0.45\textwidth]{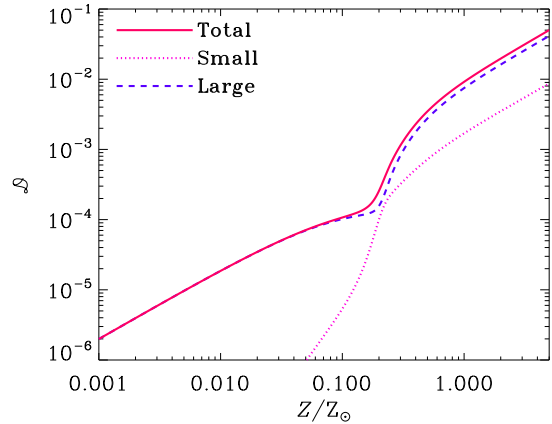}
\caption{Evolution of dust-to-gas ratio as a function of
metallicity $Z$ normalized to the solar
metallicity $Z_{\sun}=0.02$ for the fiducial case.
We show the dust-to-gas ratios for the total ($\mathcal{D}$; solid line),
the small grains ($\mathcal{D}_\mathrm{s}$; dotted line), and the
large grains ($\mathcal{D}_\mathrm{l}$; dashed line).
\label{fig:dust_metal}}
\end{figure}

We now show in Fig.\ \ref{fig:dust_metal} the evolution of
$\mathcal{D}_\mathrm{l}$, $\mathcal{D}_\mathrm{s}$, and
$\mathcal{D}(=\mathcal{D}_\mathrm{l}+\mathcal{D}_\mathrm{s})$
as a function of $Z$, calculated by
equations (\ref{eq:dDldt2}) and (\ref{eq:dDsdt2}).
At low metallicity, the large grains dominate the total dust abundance
because the stellar dust production is the dominant
dust formation mechanism.
{Since we assume a constant condensation efficiency
($f_\mathrm{in}$) of the metals in stellar ejecta,
the dust-to-gas ratio is proportional to the metallicity
($\mathcal{D}\simeq\mathcal{D}_\mathrm{l}\simeq f_\mathrm{in}Z$)
at low metallicity (see also Section \ref{subsec:lowZ}).}
The slowdown of the increase
around $Z\sim 0.1$ Z$_{\sun}$ is due to the destruction by SN
shocks. At the same time, the small grains increase because
shattering time-scale becomes short enough to produce small
grains through the collisions of the large grains. After that,
accretion
further accelerates the increase of the small grains, and
at the same time, the large grains rapidly increase because
coagulation of small grains becomes active.
{This rapid increase of both small and large grain
abundances appears around $Z\sim 0.2$ Z$_{\sun}$.}
This increase
slows down after a significant fraction of gas-phase metals
are used up.
{The increase of dust-to-gas ratio is driven by the
metal enrichment at high metallicity, where the
dust-to-metal ratio is determined by the balance between
accretion and SN destruction.
The small-to-large grain abundance ratio is determined by
the balance between shattering and coagulation.
See Section \ref{subsec:highZ} for quantitative discussions
for the high-metallicity regime.}

The evolutionary behaviours of the small and large grain components
match the full treatment of grain size distribution in
A13 (see their fig.\ 6). For the case of $\tau_\mathrm{SF}=5$ Gyr in
A13, the dust grains are predominantly large at the early stage of
galaxy evolution, while small grains are supplied
after $\sim$1\,Gyr when the metallicity reaches $\sim$0.2\,Z$_{\sun}$.
This metallicity level just matches the one at which the rapid
increase of the small grain abundance is seen in
Fig.\ \ref{fig:dust_metal}. The large-grain abundance also
increases as a result of coagulation.

\subsection{Parameter dependence}

The effects of individual processes can be investigated by changing
the parameters listed in Table \ref{tab:param}. In what follows,
we describe each process, where we only change the parameter
that regulates it, with the other parameters fixed at
the fiducial values.

\subsubsection{Dust formation in stellar ejecta}

\begin{figure}
\includegraphics[width=0.45\textwidth]{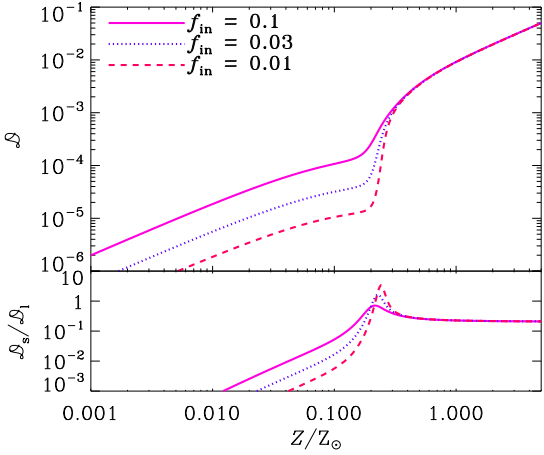}
\caption{Same as Fig.\ \ref{fig:dust_metal} but for various
dust condensation efficiencies in stellar ejecta
$f_\mathrm{in}$. In the upper window, the total dust-to-gas ratio
$\mathcal{D}$ is shown.
The solid, dotted, and dashed lines show the cases with
$f_\mathrm{in}=0.1$, 0.03, and 0.01, respectively.
In the lower window, the small-to-large grain abundance ratio is shown.
\label{fig:fin}}
\end{figure}

The amount of dust ejected by stars into the ISM is regulated
by the condensation efficiency $f_\mathrm{in}$ in our model,
in which we consider the range $f_\mathrm{in}=0.01$--0.1.
In Fig.\ \ref{fig:fin}, we show the relation between
dust-to-gas ratio and metallicity
for $f_\mathrm{in}=0.1$, 0.03, and 0.01. We also present the
small-to-large grain abundance ratio,
$\mathcal{D}_\mathrm{s}/\mathcal{D}_\mathrm{l}$.
We observe that the difference in $f_\mathrm{in}$
leads to a different dust-to-gas ratio proportional to
$f_\mathrm{in}$ at low metallicity before accretion
drastically increases the dust-to-gas ratio.
The metallicity at which accretion becomes efficient
(critical metallicity; Section \ref{subsec:crit}) is
insensitive to $f_\mathrm{in}$. After this rapid increase, all the
lines converge into the same $\mathcal{D}$--$Z$ relation,
since the stellar dust production has little influence on the
total dust mass compared with accretion at high metallicity.

As shown in Fig.\ \ref{fig:fin}, the
small-to-large grain abundance ratio,
$\mathcal{D}_\mathrm{s}/\mathcal{D}_\mathrm{l}$, is
smaller for a smaller value of $f_\mathrm{in}$ at low
metallicity, simply because
the small grain production by shattering is less efficient for
small $f_\mathrm{in}$ (i.e.\ small large-grain abundance).
Associated with the
rapid increase of $\mathcal{D}$ by accretion, there appears an epoch
at which $\mathcal{D}_\mathrm{s}/\mathcal{D}_\mathrm{l}$
reaches its peak.
At this stage, $\mathcal{D}_\mathrm{s}/\mathcal{D}_\mathrm{l}$
reaches a larger value for smaller $f_\mathrm{in}$
simply because the increase of $\mathcal{D}_\mathrm{s}$
is prominent if we normalize it to $\mathcal{D}_\mathrm{l}$,
which is smaller at this stage for smaller $f_\mathrm{in}$.
After this rapid increase, the small-to-large grain abundance
ratio converges to a constant value $\sim 0.1$, which is
independent of $f_\mathrm{in}$; this value is determined by
the balance between shattering and coagulation
(Section \ref{subsec:highZ}).

\subsubsection{Dust destruction in SN shocks}\label{subsubsec:SN}

\begin{figure}
\includegraphics[width=0.45\textwidth]{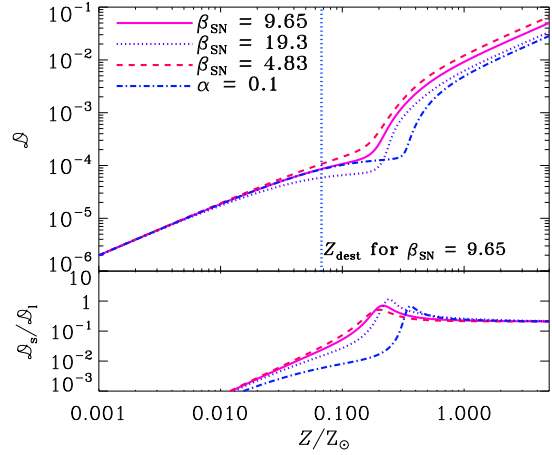}
\caption{Same as Fig.\ \ref{fig:fin} but for various
destruction efficiencies parameterized by $\beta_\mathrm{SN}$
and $\alpha$. The solid, dotted, and dashed curves show
the results with $\beta_\mathrm{SN}=9.65$, 19.3, and 4.83,
respectively ($\alpha =1$).
The dot-dashed curve represents the case with $\alpha =0.1$
(enhanced destruction of small grains) for $\beta_\mathrm{SN}=9.65$.
The vertical dotted line marks the characteristic metallicity
$Z_\mathrm{dest}$ at which the dust loss by destruction becomes
comparable to the dust supply from stars for $\beta_\mathrm{SN}=9.65$.
\label{fig:SN}}
\end{figure}

We examine the effect of dust destruction in SN shocks, i.e.\
the dependence on $\beta_\mathrm{SN}$.
In Fig.\ \ref{fig:SN}, we show the relation between dust-to-gas
ratio and
metallicity for $\beta_\mathrm{SN}=9.65$, 19.3, and 4.83
(i.e.\ the fiducial case, two times stronger destruction, and
two times weaker destruction).
The difference in $\beta_\mathrm{SN}$
starts to appear around $Z\sim 0.05$ Z$_{\sun}$, after which the
increase of $\mathcal{D}$ is more suppressed for a larger destruction
efficiency. This metallicity level is roughly estimated by
$Z_\mathrm{dest}=\mathcal{Y}_Z/\beta_\mathrm{SN}
\sim 0.067(\mathcal{Y}_Z/0.013)(\beta_\mathrm{SN}/9.65)^{-1}$ Z$_{\sun}$
(Section \ref{subsec:lowZ}; \citealt{kuo13}), and is
marked in Fig.\ \ref{fig:SN} for $\beta_\mathrm{SN}=9.65$
(and $\mathcal{Y}_Z=0.013$, which is unchanged throughout this paper).

{Just after the slight suppression of dust mass increase
around $Z_\mathrm{dest}$, the dust-to-gas ratio
(especially, the small-grain dust-to-gas ratio) rapidly
increases because of the dust mass growth by the accretion on
the small grains created by shattering. This occurs at the critical
metallicity of accretion. Above this critical metallicity,
the dust growth time-scale becomes shorter than
the destruction time-scale so that accretion overcomes the
dust destruction effect.}
The dust-to-gas ratio after the rapid growth is also smaller
for larger $\beta_\mathrm{SN}$ since the dust abundance
(more precisely, the dust-to-metal ratio) is
determined by the balance between accretion and destruction
at high metallicity (Section \ref{subsec:highZ}).
{The small-to-large grain abundance ratio, on the other
hand, is determined by the balance between shattering and
coagulation at high metallicity (Section \ref{subsec:highZ}) so that
it is insensitive to $\beta_\mathrm{SN}$ at $Z\ga 0.7$ Z$_{\sun}$.
Note that, although dust destruction and accretion
indeed have a consequence on the relative abundance of small and large
grains, the
dominant mechanism is shattering and coagulation, which, however,
do not change the total dust mass.}

We also examine the case of $\alpha <1$, which corresponds
to enhanced destruction of small grains compared with large
grains. In Fig.\ \ref{fig:SN}, we observe that the small value
of $\alpha$ delays the dust mass growth by accretion.
This is because accretion is suppressed because more small
grains are destroyed. Nevertheless, because of the strong metallicity
dependence of accretion,
accretion still becomes efficient enough around
$Z=0.3$ Z$_{\sun}$ to boost
$\mathcal{D}_\mathrm{s}/\mathcal{D}_\mathrm{l}$ and $\mathcal{D}$.

\subsubsection{Accretion}

\begin{figure}
\includegraphics[width=0.45\textwidth]{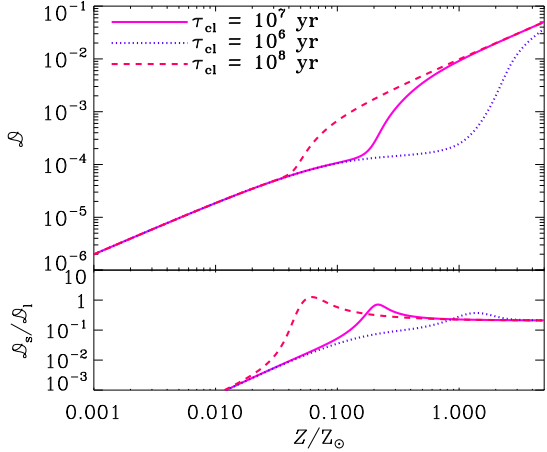}
\caption{Same as Fig.\ \ref{fig:fin} but for various
dense cloud lifetimes ($\tau_\mathrm{cl}$), which regulate the efficiency of
accretion.
The solid, dotted, and dashed lines show
the results with $\tau_\mathrm{cl}=10^7$, $10^6$, and
$10^8$ yr, respectively.
\label{fig:acc}}
\end{figure}

As explained in Section \ref{subsec:parameter}, we vary
$\tau_\mathrm{cl}$ to change the efficiency of accretion.
In Fig.\ \ref{fig:acc}, we show the relation between
dust-to-gas ratio and metallicity for various $\tau_\mathrm{cl}$.
As expected, the rapid increase of $\mathcal{D}$ by accretion
occurs earlier for
longer $\tau_\mathrm{cl}$ (i.e.\ more efficient accretion).
{As assumed in this model based on \citet{hirashita12}
(see also Section \ref{subsec:review}), accretion increases the
abundance of the small grains. Therefore, the rapid rise of
$\mathcal{D}$ is associated with the increase of
$\mathcal{D}_\mathrm{s}/\mathcal{D}_\mathrm{l}$.}
After this rapid growth is settled,
the $\mathcal{D}$--$Z$ and
$\mathcal{D}_\mathrm{s}/\mathcal{D}_\mathrm{l}$--$Z$ relations
converge to the curves independent of
$\tau_\mathrm{cl}$ at high metallicity. In fact, at
high metallicity, grain growth by accretion is saturated,
and $\mathcal{B}\simeq (Z-\mathcal{D})/\mathcal{D}_\mathrm{s}$
(equation \ref{eq:beta}; this means that all the gas phase
metals are accreted onto the small grains in dense clouds),
which is independent of $\tau_\mathrm{cl}$.

\subsubsection{Shattering}

\begin{figure}
\includegraphics[width=0.45\textwidth]{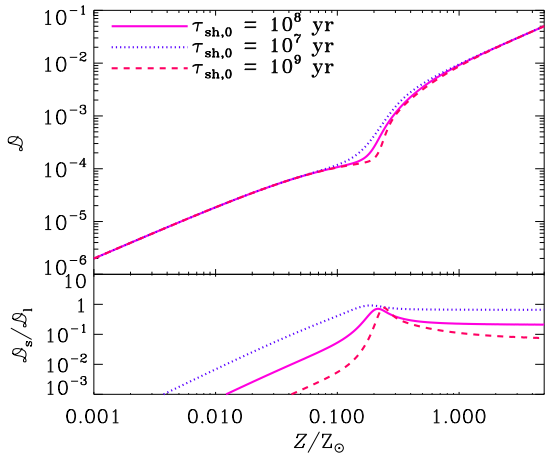}
\caption{Same as Fig.\ \ref{fig:fin} but for various
shattering time-scales at the Milky Way dust-to-gas ratio
($\tau_\mathrm{sh,0}$).
The solid, dotted, and dashed lines show
the results with $\tau_\mathrm{sh,0}=10^8$, $10^7$, and
$10^9$~yr, respectively.
\label{fig:shat}}
\end{figure}

We also change the shattering time-scale, which is
scaled with the value $\tau_\mathrm{sh,0}$ at
the Milky Way dust-to-gas ratio.
{}From Fig.\ \ref{fig:shat}, we observe that the
small-to-large grain abundance ratio is largely affected by
the shattering time-scale. This ratio is roughly proportional
to $\tau_\mathrm{sh,0}$ except at the epoch when
accretion rapidly raise the {small-grain abundance}.
Furthermore, the metallicity at which the accretion increase
the dust-to-gas ratio is smaller for a shorter shattering time-scale.
However, once accretion starts to increase the dust mass,
it rapidly boosts the small-to-large grain abundance ratio,
so that the resulting track on the $\mathcal{D}$--$Z$ plane is
not very sensitive to shattering.
Shattering does not directly change the dust abundance, but
it plays an important role in the total dust
abundance
{through its capability of producing a large number of} small grains,
whose growth by accretion is of fundamental importance in
the dust mass increase (see also \citealt{kuo12} and A13).

\subsubsection{Coagulation}

\begin{figure}
\includegraphics[width=0.45\textwidth]{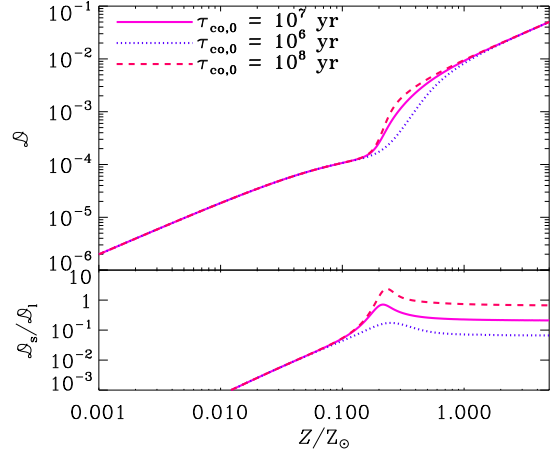}
\caption{Same as Fig.\ \ref{fig:fin} but for various
coagulation time-scales.
The solid, dotted, and dashed lines show
the results with $\tau_\mathrm{co,0}=10^7$, $10^6$, and
$10^8$ yr, respectively.
\label{fig:coag}}
\end{figure}

For the variation of the coagulation time-scale, we
show the results in Fig.\ \ref{fig:coag}. We observe that
the small-to-large grain abundance ratio is
independent of $\tau_\mathrm{co,0}$ at low
metallicity ($Z<0.03$ Z$_{\sun}$) while it is larger
for larger $\tau_\mathrm{co}$. This is because
coagulation has little influence on the dust evolution
at low metallicity and the small grains are efficiently
converted into large grains after the dust-to-gas ratio
has become large enough for coagulation to occur.
The metallicity at which accretion rapidly increases the
dust-to-gas ratio is not sensitive to the coagulation
time-scale, although the growth is significantly
suppressed for the case of the shortest
$\tau_\mathrm{co,0}$, since a major part of the small
grains are converted into the large grains before
they accrete the gas-phase metals. The final
dust-to-gas ratio at high metallicity is insensitive to
the coagulation time-scale.

\subsubsection{Star formation time-scale}

\begin{figure}
\includegraphics[width=0.45\textwidth]{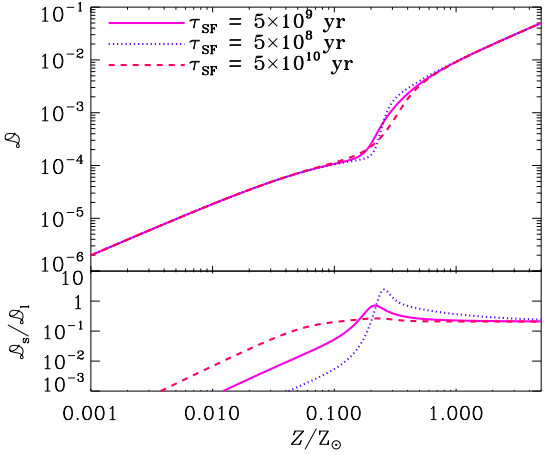}
\caption{Same as Fig.\ \ref{fig:fin} but for various
star formation time-scales.
The solid, dotted, and dashed lines show
the results with $\tau_\mathrm{SF}=5\times 10^9$,
$5\times 10^8$, and $5\times 10^{10}$ yr, respectively.
\label{fig:sf}}
\end{figure}

We also change the star formation time-scale, $\tau_\mathrm{SF}$,
which governs the metal-enrichment time-scale of the galaxy.
This time-scale affects shattering and coagulation, since
$\beta_\mathrm{sh}$ ($\beta_\mathrm{co}$) is determined
by the ratio of $\tau_\mathrm{SF}$ to $\tau_\mathrm{sh}$
($\tau_\mathrm{co}$). In Fig.\ \ref{fig:sf}, we show the
results for $\tau_\mathrm{SF}=5\times 10^8$, $5\times 10^9$,
and $5\times 10^{10}$ yr. The $\mathcal{D}$--$Z$ relation is hardly affected
by $\tau_\mathrm{SF}$. In contrast,
\citet{inoue03} concluded that $\mathcal{D}$--$Z$ is largely
affected by $\tau_\mathrm{SF}$. This is because \citet{inoue03}
moved $\tau_\mathrm{SF}$ freely, while we assume
the proportionality between $\tau_\mathrm{SF}$ and
$\tau_\mathrm{acc}$ through equations (\ref{eq:dmdt_acc})
and (\ref{eq:SFR}) since
both star formation and accretion occur in the dense ISM.

As observed in the bottom panel {of Fig.\ \ref{fig:sf}},
the small-to-large grain abundance
ratio is largely varied by the change of $\tau_\mathrm{SF}$
{especially at low metallicity. At low metallicity,
the small-grain abundance is governed not only by
the abundance of the large grains, from which the small grains
are produced by shattering, but also by how much shattering can
occur in the metal-enrichment time-scale (i.e.,
$\tau_\mathrm{SF}/\tau_\mathrm{sh,0}$). Therefore, if we
take the small-to-large grain abundance ratio, the factor
$\tau_\mathrm{SF}/\tau_\mathrm{sh,0}$ remains
(see equation \ref{eq:DsDl} for more quantification).}
Thus, $\mathcal{D}_\mathrm{s}/\mathcal{D}_\mathrm{l}$
is larger for longer $\tau_\mathrm{SF}$ at low metallicity.
The `overshoot' of
$\mathcal{D}_\mathrm{s}/\mathcal{D}_\mathrm{l}$ around
0.2~Z$_{\sun}$ does not appear for the longest $\tau_\mathrm{SF}$,
since the equilibrium between shattering and coagulation is
achieved before the metallicity reaches that value. Indeed,
Fig.\ \ref{fig:sf} shows that $\mathcal{D}_\mathrm{s}/\mathcal{D}_\mathrm{l}$
converges to a constant value determined by the equilibrium between
shattering and coagulation (see Section~\ref{subsec:highZ})
even before accretion rapidly
raise the dust-to-gas ratio {for the longest $\tau_\mathrm{SF}$)}.
The `overshoot' is the most prominent for the shortest
$\tau_\mathrm{SF}$. All the three cases finally converge to the same
values of $\mathcal{D}_\mathrm{s}/\mathcal{D}_\mathrm{l}$ determined
by the equilibrium between shattering and coagulation.

\subsection{Monte Carlo plots}\label{subsec:monte}

To summarize the variation of the $\mathcal{D}$--$Z$ and
$\mathcal{D}_\mathrm{s}/\mathcal{D}_\mathrm{l}$--$Z$ relations,
we perform a Monte Carlo simulation, varying the parameters
within the ranges examined above (Table \ref{tab:param}).
Such a Monte Carlo
approach to predict the variation of the evolutionary
tracks in the relation between dust-to-gas ratio and metallicity has already
been performed by \citet{mattsson14}. We apply their method to
our formulation and set of parameters. For simplicity,
we assume a uniform distribution for the logarithm of each
parameter in the range shown in Table \ref{tab:param}.
For each set of parameter values chosen, we calculate
the evolutionary track on the $\mathcal{D}$--$Z$ and
$\mathcal{D}_\mathrm{s}/\mathcal{D}_\mathrm{l}$--$Z$ plane.
By repeating this for 100,000 random choices of the parameter
set, we obtain the density distribution of evolutionary tracks.
This implicitly assume that the metallicity distribution is
uniform. In other words, the slice at each metallicity
gives the probabilities of  $\mathcal{D}$
and $\mathcal{D}_\mathrm{s}/\mathcal{D}_\mathrm{l}$ at the
metallicity, $P_1(\mathcal{D}|Z)\,\mathrm{d}\log\mathcal{D}$ and
$P_2(\mathcal{D}_\mathrm{s}/\mathcal{D}_\mathrm{l}|Z)\,
\mathrm{d}(\mathcal{D}_\mathrm{s}/\mathcal{D}_\mathrm{l})$, respectively.

\begin{figure}
\includegraphics[width=0.45\textwidth]{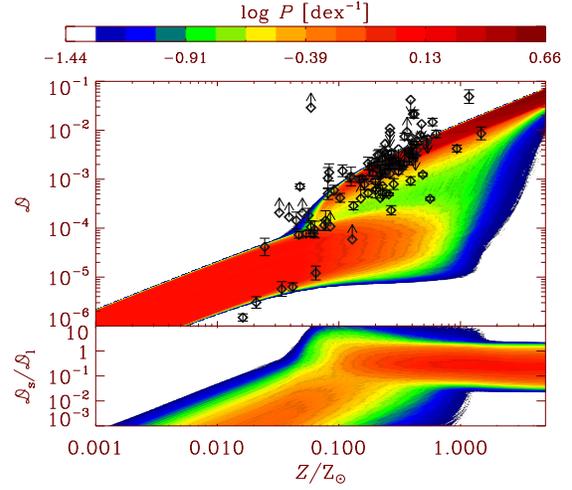}
\caption{Probability distribution on the $\mathcal{D}$--$Z$
and $\mathcal{D}_\mathrm{s}/\mathcal{D}_\mathrm{l}$--$Z$ planes
for the Monte Carlo simulation with randomly selected values of
the parameters. The probability $P$ ($=P_1$ in the upper panel
and $P_2$ in the lower panel) at each
metallicity is shown with the color maps. The scale is
shown in the colour bar above the figures. For reference,
we also plot the observational data of nearby galaxies
taken from \citet{remy14} (diamonds with error bars).
The arrows show the
upper or lower limits depending on the direction.
\label{fig:random}}
\end{figure}

In Fig.\ \ref{fig:random}, we show the colour maps of
$P_1(\mathcal{D}|Z)$ and
$P_2(\mathcal{D}_\mathrm{s}/\mathcal{D}_\mathrm{l}|Z)$.
For the $\mathcal{D}$--$Z$ relation, $P_1$ is almost constant
at low metallicity
simply because we assume uniform distribution for
$\log f_\mathrm{in}$, which is the only factor affecting the
$\mathcal{D}$--$Z$ relation.
At $Z\ga 0.1$ Z$_{\sun}$, accretion becomes the dominant
mechanism of dust mass increase, producing the peak of
$P_1$ at large dust-to-gas ratio. At
$0.1\la Z/\mathrm{Z}_{\sun}\la 1$, the distribution of
dust-to-gas ratio is bimodal with different dust production
mechanisms: the upper/lower peak is composed of the objects in which
accretion/stellar ejecta is the main dust production mechanism.
Since the dust mass increase by accretion occurs rapidly,
these two branches separate clearly, producing the bimodality.

The distribution of small-to-large grain abundance ratio in
Fig.~\ref{fig:random} (bottom panel) shows a large
dispersion at low
metallicity, reflecting the variations of $f_\mathrm{in}$
(Fig.\ \ref{fig:fin}) and
$\tau_\mathrm{sh,0}/\tau_\mathrm{SF}$ (Figs.\ \ref{fig:shat}
and \ref{fig:sf})
(see also Section \ref{subsec:lowZ}).
The overshoot of the small-to-large grain abundance ratio
is clearly seen around $Z\sim 0.2$ Z$_{\sun}$. At high
metallicity the small-to-large grain abundance ratio converges
to the values determined by the shattering and coagulation
time-scales (Section \ref{subsec:highZ}); thus, the dispersion of
$\mathcal{D}_\mathrm{s}/\mathcal{D}_\mathrm{l}$ at high metallicity
reflects the
dispersions of $\tau_\mathrm{sh,0}$ and $\tau_\mathrm{co,0}$.

For reference, we also plot the observational data of
nearby galaxies in Fig.\ \ref{fig:random}. The data are
taken from \citet{remy14} \citep[see also][]{remy13}.
They combined three different samples:
the Dwarf Galaxy Survey \citep{madden13},
the KINGFISH survey \citep{kennicutt11} and a subsample of
\citet{galametz11}. They derived the dust masses by fitting the
SED with the model presented in \citet{galliano11}, and took
the atomic and molecular masses in the literature.
For the CO-to-H$_2$ conversion factor, we adopt their
metallicity-dependent case, but the results and discussions
does not change even if we adopt the Milky Way conversion
factor. For the solar metallicity, we adopt
$12+\log (\mathrm{O/H})=8.93$ \citep{anders89} to be
consistent with the normalization used for the models.

In Fig.\ \ref{fig:random}, we observe that the distribution
of the data is broadly consistent with the colour map.
The trend from the low dust-to-gas ratio at low metallicity
to the high dust-to-gas ratio at high
metallicity is correctly reproduced.
In particular,
the region with the highest probabilities on the $\mathcal{D}$--$Z$
diagram at $0.1\la Z/\mathrm{Z}_{\sun}\la 1$, where the main dust
formation mechanism is accretion, matches the
concentration of the observational data. This supports the
necessity of accretion for this metallicity range
{(see also the models in \citealt{remy14} and
\citet{zhukovska14}, and figure 5 in
\citealt{debennassuti14} for comparisons with the same data)}.
A few low-$\mathcal{D}$ objects in this metallicity range are
consistent with the lower
branch (secondary peak) of $P_1$ in which the
dust is predominantly supplied from stars.
At lower metallicities, the stellar dust production explains
the data well, although the number
of low-metallicity objects is still small.
Considering the simplicity of our models, we judge
that the model and the choice of the parameter ranges are
successful.

\section{Discussion}\label{sec:discussion}

We analyze the behaviours of
$\mathcal{D}$ and $\mathcal{D}_\mathrm{s}/\mathcal{D}_\mathrm{l}$
at various metallicities.

\subsection{Evolution at low metallicity}
\label{subsec:lowZ}

At low metallicity,
if we keep the terms up to the first order
for $Z$ in equation (\ref{eq:dDldt2}), we obtain for the large grains
\begin{eqnarray}
\mathcal{Y}_Z\frac{\mathrm{d}\mathcal{D}_\mathrm{l}}{\mathrm{d}Z}
\simeq f_\mathrm{in}(\mathcal{R}Z+\mathcal{Y}_Z)-(\beta_\mathrm{SN}+
\mathcal{R})\mathcal{D}_\mathrm{l}.\label{eq:dDldZ_lowZ}
\end{eqnarray}
The zeroth order indicates the solution
$\mathcal{D}_\mathrm{l}\simeq f_\mathrm{in}Z$. This means that
only the stellar dust contributes to the dust enrichment
at low metallicity. If we put this solution on the right hand
size in equation (\ref{eq:dDldZ_lowZ}), we obtain
$\mathrm{d}\mathcal{D}_\mathrm{l}/\mathrm{d}Z\simeq f_\mathrm{in}
-\beta_\mathrm{SN}f_\mathrm{in}Z/\mathcal{Y}_Z$, where the second
term indicates the destruction in SN shocks. Comparing the first
and second terms on the right-hand side of this equation, we
find that the dust destruction almost cancels the dust formation
by stars at a characteristic metallicity, $Z_\mathrm{dest}$:
$Z\sim Z_\mathrm{dest}\equiv\mathcal{Y}_Z/\beta_\mathrm{SN}\sim
0.067(\mathcal{Y}_Z/0.013)(\beta_\mathrm{SN}/9.65)^{-1}$ Z$_{\sun}$,
the corresponding dust-to-gas
ratio being $\mathcal{D}_\mathrm{l}\sim f_\mathrm{in}Z_\mathrm{dest}
\sim 1.3\times 10^{-4}(\mathcal{Y}_Z/0.013)
(\beta_\mathrm{SN}/9.65)^{-1}(f_\mathrm{in}/0.1)$.
Indeed, $Z=Z_\mathrm{dest}$, marked in Fig.\ \ref{fig:SN},
roughly corresponds to the
metallicity level at which the increase of dust-to-gas
ratio slows down significantly.

For the small grains, if we only take the lowest order
for $Z$, the second-order term representing shattering
remains:
\begin{eqnarray}
\mathcal{Y}_Z\frac{\mathrm{d}\mathcal{D}_\mathrm{s}}{\mathrm{d}Z}
=\beta_\mathrm{sh}\mathcal{D}_\mathrm{l}=
\frac{\tau_\mathrm{SF}}{\tau_\mathrm{sh,0}}
\frac{1}{\mathcal{D}_\mathrm{MW,s}}(f_\mathrm{in}Z)^2,
\end{eqnarray}
where we used the solution at low metallicity,
$\mathcal{D}_\mathrm{l}=f_\mathrm{in}Z$. Solving this,
and using the low-metallicity solution for $\mathcal{D}_\mathrm{l}$,
we obtain the expression valid for low metallicities:
\begin{eqnarray}
\frac{\mathcal{D}_\mathrm{s}}{\mathcal{D}_\mathrm{l}} & \simeq &
\frac{f_\mathrm{in}Z^2}{3\mathcal{Y}_Z\mathcal{D}_\mathrm{MW,l}}
\left(\frac{\tau_\mathrm{SF}}{\tau_\mathrm{sh,0}}\right)\nonumber\\
& = & 0.073\left(\frac{f_\mathrm{in}}{0.1}\right)
\left(\frac{\tau_\mathrm{SF}}{5\times 10^9~\mathrm{yr}}\right)
\left(\frac{\tau_\mathrm{sh,0}}{10^8~\mathrm{yr}}\right)^{-1}\nonumber\\
& & \times\left(\frac{Z}{0.1~\mathrm{Z}_{\sun}}\right)^2,\label{eq:DsDl}
\end{eqnarray}
where we fix $\mathcal{Y}_Z=0.013$ and $\mathcal{D}_\mathrm{MW,l}=0.007$.
This nicely explains the behaviour 
and parameter dependence of the small-to-large grain abundance
ratio at low metallicity (note especially the dependence on
$f_\mathrm{in}$, $\tau_\mathrm{sh,0}$, and $\tau_\mathrm{SF}$ shown in
Figs.\ \ref{fig:fin}, \ref{fig:shat}, and \ref{fig:sf}, respectively).

\subsection{Evolution at high metallicity}
\label{subsec:highZ}

At high metallicity, the dust-to-gas ratio increases
in proportion to the metallicity, which means that the
dust-to-metal ratio $\mathcal{D}/Z$ converges to a
constant value. We also observe that
 the small-to-large
grain abundance ratio converges to a constant value at high
metallicity. These characteristics are analyzed by taking
the differentials of $\mathcal{D}_\mathrm{s}/Z$ and
$\mathcal{D}_\mathrm{l}/Z$ as follows.

At high metallicity, we can neglect the dust supply from
stars, and the dust content is determined by interstellar
processing (accretion, destruction, shattering and coagulation).
By keeping those terms, the differentials of $\mathcal{D}_\mathrm{s}/Z$ and
$\mathcal{D}_\mathrm{l}/Z$ are expressed, using equations
(\ref{eq:dDldt2}) and (\ref{eq:dDsdt2}) as
\begin{eqnarray}
\mathcal{Y}_Z\frac{\mathrm{d}}{\mathrm{d}Z}\left(
\frac{\mathcal{D}_\mathrm{l}}{Z}\right) & \hspace{-2mm}\simeq &
\hspace{-2mm}\frac{1}{Z}
[\beta_\mathrm{co}\mathcal{D}_\mathrm{s}-(\beta_\mathrm{SN}+
\beta_\mathrm{sh})\mathcal{D}_\mathrm{l}],\label{eq:dust_metal_l}\\
\mathcal{Y}_Z\frac{\mathrm{d}}{\mathrm{d}Z}\left(
\frac{\mathcal{D}_\mathrm{s}}{Z}\right) & \hspace{-2mm}\simeq &
\hspace{-2mm}\frac{1}{Z}
\left[\beta_\mathrm{sh}\mathcal{D}_\mathrm{l}-\left(
\frac{\beta_\mathrm{SN}}{\alpha}+\beta_\mathrm{co}-
\beta_\mathrm{acc}\right)\mathcal{D}_\mathrm{s}\right] .\nonumber\\
\label{eq:dust_metal_s}
\end{eqnarray}
The terms with $\beta_\mathrm{co}$ and $\beta_\mathrm{sh}$ grow
as metallicity (or dust-to-gas ratio) increases; thus, they
eventually become {dominant}. This indicates that
coagulation and shattering balance at high metallicity.
In other words,
$\beta_\mathrm{co}\mathcal{D}_\mathrm{s}-\beta_\mathrm{sh}
\mathcal{D}_\mathrm{l}\simeq 0$, leading to the expectation that
the small-to-large grain abundance ratio converge to
$\mathcal{D}_\mathrm{s}/\mathcal{D}_\mathrm{l}\simeq
\beta_\mathrm{sh}/\beta_\mathrm{co}=\tau_\mathrm{co}/\tau_\mathrm{sh}$.
Using equations (\ref{eq:tau_shat}) and (\ref{eq:tau_coag}),
we obtain
\begin{eqnarray}
\frac{\mathcal{D}_\mathrm{s}}{\mathcal{D}_\mathrm{l}}
\simeq\left(
\frac{\mathcal{D}_\mathrm{MW,s}}{\mathcal{D}_\mathrm{MW,l}}\,
\frac{\tau_\mathrm{co,0}}{\tau_\mathrm{sh,0}}\right)^{1/2}
\label{eq:dgs_dgl}
\end{eqnarray}
for the high-metallicity regime. If we put the fiducial
values, we obtain
$\mathcal{D}_\mathrm{s}/\mathcal{D}_\mathrm{l}\simeq 0.21$,
explaining the value at high metallicity very well.
Equation (\ref{eq:dgs_dgl}) also gives correct estimates
for other values of $\tau_\mathrm{sh,0}$ and $\tau_\mathrm{co,0}$.

If we add equations (\ref{eq:dust_metal_l}) and
(\ref{eq:dust_metal_s}), we obtain the equation for
the total dust-to-metal ratio:
\begin{eqnarray}
\mathcal{Y}_Z\frac{\mathrm{d}}{\mathrm{dZ}}\left(
\frac{\mathcal{D}}{Z}\right)
\simeq\frac{1}{Z}\left[ -\beta_\mathrm{SN}\delta\mathcal{D}
+\beta_\mathrm{acc}\mathcal{D}_\mathrm{s}\right] ,
\end{eqnarray}
where
\begin{eqnarray}
\delta\equiv 1+\left(\frac{1}{\alpha}-1\right)
\frac{\mathcal{D}_\mathrm{s}}{\mathcal{D}}.
\end{eqnarray}
If $\alpha =1$, $\delta =1$. For high
metallicity,
$\beta_\mathrm{acc}=\beta /\epsilon\sim (Z-\mathcal{D})/
(\epsilon\mathcal{D}_\mathrm{s})$
($y\to\infty$, which corresponds to the most efficient
accretion, in equation \ref{eq:beta}), so
that we obtain
\begin{eqnarray}
\frac{\mathcal{D}}{Z}\simeq
\frac{1}{\epsilon\beta_\mathrm{SN}\delta +1}.
\end{eqnarray}
For $\alpha =1$ ($\delta =1$), this gives the same result
as \citet{hirashita11}. For the fiducial case,
$\mathcal{D}/Z\sim 0.5$, which explains the
dust-to-metal ratio at high metallicity.
This is consistent with the metal depletion in the
Milky Way \citep[e.g.][]{kimura03}.
If the small grains are more
easily destroyed (i.e.\ $\alpha <1$, so $\delta >1$), the dust-to-metal
ratio is suppressed, which reflects the fact that
that accretion is also suppressed because of the
destruction of the small grains. This explains the
suppression of $\mathcal{D}$ in the case of $\alpha =0.1$
compared with the other
cases in Fig.\ \ref{fig:SN}.

\subsection{Critical metallicity for accretion}\label{subsec:crit}

The most remarkable increase of dust-to-gas ratio is
induced by accretion. Because of its metallicity dependence,
accretion occurs after the ISM is enriched with metals.
The metallicity level at which
accretion starts to increase the dust-to-gas ratio significantly
is referred to as the critical metallicity for accretion
\citep{inoue11,asano13b}. Following \citet{hirashita11},
we estimate the critical metallicity with
$\beta_\mathrm{SN}\mathcal{D}=\beta_\mathrm{acc}\mathcal{D}_\mathrm{s}$,
which means that accretion starts to increase the dust-to-gas
ratio more than SN destruction decreases it
(equation \ref{eq:dg_metal}; note that $\mathcal{R}\ll\beta_\mathrm{SN}$
and that $\beta_\mathrm{SN}'=\beta_\mathrm{SN}$ for $\alpha =1$).
We observe from Figs.\ \ref{fig:fin}--\ref{fig:coag} that
$\mathcal{D}_\mathrm{s}/\mathcal{D}_\mathrm{l}$ approaches
$\sim 1$ except for the case of strong coagulation
(i.e.\ $\tau_\mathrm{co,0}\la 10^6$ yr).
Thus, we adopt $\mathcal{D}_\mathrm{s}\simeq\mathcal{D}/2$,
obtaining $\beta_\mathrm{acc}/2 \simeq\beta_\mathrm{SN}$ at
the critical metallicity.

\begin{figure}
\includegraphics[width=0.45\textwidth]{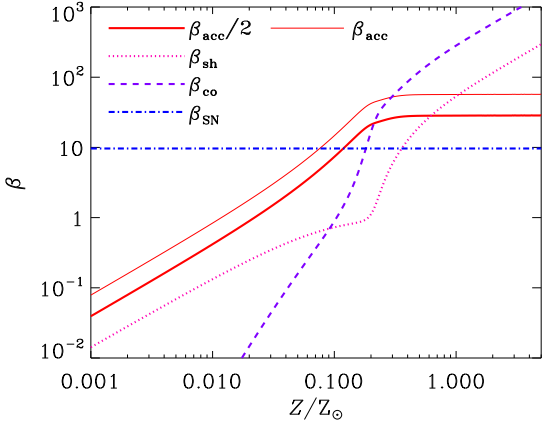}
\caption{For comparison among the efficiencies of various
processes, we plot $\beta_\mathrm{acc}/2$, $\beta_\mathrm{acc}$, 
$\beta_\mathrm{sh}$,
$\beta_\mathrm{co}$, and $\beta_\mathrm{SN}$ as a function of
metallicity (thick solid, thin solid, dotted, dashed, and dot-dashed
lines, respectively) for the fiducial case.
\label{fig:beta}}
\end{figure}

In Fig.~\ref{fig:beta}, we compare $\beta_\mathrm{acc}/2$ and
$\beta_\mathrm{SN}$ for the fiducial case. As the metallicity increases,
$\beta_\mathrm{acc}$ grows. The metallicity at the crossover
of $\beta_\mathrm{SN}$ and $\beta_\mathrm{acc}/2$ is the critical
metallicity, which is 0.12 Z$_{\sun}$ in this case. Around this
metallicity, the rapid increase of
dust-to-gas ratio indeed occurs in Fig.\ \ref{fig:dust_metal}.

In Fig.\ \ref{fig:beta}, we also show $\beta_\mathrm{sh}$ and
$\beta_\mathrm{co}$, which are proportional to
$\mathcal{D}_\mathrm{l}$ and $\mathcal{D}_\mathrm{s}$,
respectively. We observe that $\beta_\mathrm{co}$ becomes
comparable to $\beta_\mathrm{acc}$ just above the
critical metallicity because the rapid increase of the
small grain abundance by accretion enhances the coagulation rate.
The metallicity at the maximum of
$\mathcal{D}_\mathrm{s}/\mathcal{D}_\mathrm{l}$ roughly
corresponds to the point where
$\beta_\mathrm{co}\simeq\beta_\mathrm{acc}$.
After that, coagulation becomes efficient and convert the small
grains into the large grains. Thus, it is predicted that
at medium metallicity galaxies experience an epoch at which the abundance of
small grains is enhanced.

It is possible to make a rough estimate of the maximum
$\mathcal{D}_\mathrm{s}/\mathcal{D}_\mathrm{l}$. As discussed
above, $\beta_\mathrm{co}\sim\beta_\mathrm{acc}$
is satisfied when
$\mathcal{D}_\mathrm{s}/\mathcal{D}_\mathrm{l}$ reaches
its maximum. As shown in Fig.\ \ref{fig:beta},
$\beta_\mathrm{acc}$ saturates above $\beta_\mathrm{SN}$.
Thus, we roughly adopt $\beta_\mathrm{acc}\sim 3\beta_\mathrm{SN}$
for analytic simplicity.
Using the definition $\beta_\mathrm{co}=\tau_\mathrm{SF}/\tau_\mathrm{co}$
and equation~(\ref{eq:tau_shat}), we obtain
$\mathcal{D}_\mathrm{s}\sim 3\beta_\mathrm{SN}
(\tau_\mathrm{co,0}/\tau_\mathrm{SF})\mathcal{D}_\mathrm{MW,s}$.
For the large grains, just before the rapid increase by
coagulation, $\mathcal{D}_\mathrm{l}$ is assumed to be
determined by the
balance between the formation in stellar ejecta and the
destruction by SNe, so that
$\mathcal{D}_\mathrm{l}\sim f_\mathrm{in}Z_\mathrm{dest}
\sim 1.3\times 10^{-4}(\mathcal{Y}_Z/0.013)
(\beta_\mathrm{SN}/9.65)^{-1}(f_\mathrm{in}/0.1)$
(Section \ref{subsec:lowZ}). Thus,
the maximum $\mathcal{D}_\mathrm{s}/\mathcal{D}_\mathrm{l}$
is estimated as
\begin{eqnarray}
\left(\frac{\mathcal{D}_\mathrm{s}}{\mathcal{D}_\mathrm{l}}\right)_\mathrm{max}
& \sim & \left(
\frac{3\beta_\mathrm{SN}\mathcal{D}_\mathrm{MW,s}}{f_\mathrm{in}Z_\mathrm{dest}}
\right)
\left(\frac{\tau_\mathrm{co,0}}{\tau_\mathrm{SF}}\right)\nonumber\\
& \sim & 1.3\left(\frac{\beta_\mathrm{SN}}{9.65}\right)^2
\left(\frac{\tau_\mathrm{co,0}}{10^7~\mathrm{yr}}\right)
\left(\frac{\tau_\mathrm{SF}}{5\times 10^9~\mathrm{yr}}\right)^{-1}\nonumber\\
& & \times\left(\frac{f_\mathrm{in}}{0.1}\right)^{-1}
\left(\frac{\mathcal{Y}_Z}{0.013}\right)^{-1}
\left(\frac{\mathcal{D}_\mathrm{MW,s}}{0.003}\right) .
\end{eqnarray}
This roughly explains the value of the maximum small-to-large
grain abundance ratio and its dependence on $f_\mathrm{in}$,
$\beta_\mathrm{SN}$, $\tau_\mathrm{co,0}$, and $\tau_\mathrm{SF}$
shown in
Figs.\ \ref{fig:fin}, \ref{fig:SN}, \ref{fig:coag}, and \ref{fig:sf},
respectively. We also find in the above expression that
the maximum $\mathcal{D}_\mathrm{s}/\mathcal{D}_\mathrm{l}$
is independent of shattering, which is consistent with the
constant maximum $\mathcal{D}_\mathrm{s}/\mathcal{D}_\mathrm{l}$
achieved in Fig.\ \ref{fig:shat}.

\subsection{Evolution of grain size distribution}

Our simple two-size approach successfully
catches the characteristics of grain size evolution
in A13's full calculation of grain size distribution.
In the early stage of galaxy evolution when the metallicity
is much smaller than the critical metallicity of accretion,
the dust is predominantly enriched by stellar ejecta.
In this stage, the large grains dominate the
dust content. The only path of small grain formation is
shattering, which becomes more and more efficient as
the system is enriched with the large grains. At a
metallicity level referred to as the critical metallicity,
the abundance of small grains formed by shattering
becomes large enough to activate the grain
growth by accretion, which contributes to the rapid
increase of dust-to-gas ratio. Associated with this phase,
the small-to-large grain abundance ratio becomes the largest
in the entire life of the galaxy.
After the increase of dust-to-gas ratio by accretion is
saturated, the small-to-large grain abundance ratio is
governed by the balance between shattering and
coagulation.

Equation (\ref{eq:dgs_dgl}), valid for high-metallicity
galaxies, can be used to constrain the
ratio of the coagulation time-scale to the shattering time-scale.
In particular, in the Milky Way, the self-consistent
small-to-large grain abundance ratio can be estimated as
$\mathcal{D}_\mathrm{MW,s}/\mathcal{D}_\mathrm{MW,l}
\simeq\tau_\mathrm{co,0}/\tau_\mathrm{sh,0}$.
Using $\mathcal{D}_\mathrm{MW,s}/\mathcal{D}_\mathrm{MW,l}$
estimated in Section \ref{subsec:evol}
based on the MRN grain size distribution, we obtain
$\tau_\mathrm{co,0}/\tau_\mathrm{sh,0}\simeq 0.43$ in
the Milky Way.
The ratio should depend on the mass fractions of various ISM
phases \citep{seok14}. Although it is difficult to determine the
absolute values of $\tau_\mathrm{sh,0}$ and $\tau_\mathrm{co,0}$,
putting a constraint on the ratio of these two
time-scales is a big step to understand
the evolution of grain size distribution in galaxies.

\section{Future prospects}\label{sec:prospect}

\subsection{Implementation in complex galaxy evolution models}

As mentioned in the Introduction, one of the major purposes of
developing computationally light models for the evolution of
grain size distribution is to implement it in complex
galaxy evolution models. It is extremely difficult to
calculate the full evolution of grain size distribution,
since it is simply expected that the computational time
increases in proportion to the number of grain size bins.
Many galaxy evolution models
already treat metal enrichment so that it is rather
straightforward
to calculate equations (\ref{eq:dDldt2}) and (\ref{eq:dDsdt2})
in such models. We could apply them to each galaxy or
each SPH particle. Variation of $\tau_\mathrm{co,0}$ and
$\tau_\mathrm{sh,0}$ can also be included based on
the mixture of various ISM phases \citep{seok14} or
based on the density and temperature of each SPH particle
\citep{hirashita09}, as straightforward extensions.

As a trade-off of the simplicity, there is of course a demerit of
using such a simple model
as developed in this paper. Since it does not treat the full
information of the grain size distribution, we need to assume
a functional form of grain size distribution
in calculating observational dust properties. Representative
observational quantities related to dust are extinction
curves and infrared dust emission SEDs. In the following
subsections, we discuss how to calculate these quantities
using the two-size model, in order to provide future prospects.
The real implementation of our models to galaxy evolution models
will be done in future work.

Here we propose to use a `modified-lognormal function' for
the grain size distributions of the small and large grains
(the grain size distribution is defined so that $n_i(a)\,\mathrm{d}a$
is the number of grains per hydrogen nucleus in the
range of grain radii between $a$ and $a+\mathrm{d}a$):
\begin{eqnarray}
n_i(a)=\frac{C_i}{a^4}\exp\left\{ -
\frac{[\ln (a/a_{0,i})]^2}{2\sigma^2}\right\} ,\label{eq:lognormal}
\end{eqnarray}
where subscript $i$ indicates the small ($i=\mathrm{s}$) or large
($i=\mathrm{l}$) grain component,
$C_i$ is the normalization constant, and $a_{0,i}$ and
$\sigma$ are the central grain radius and the standard deviation
of the lognormal distribution, respectively.
The functional form
is adopted so that the mass distribution $\propto a^3n_i(a)$
is lognormal. We adopt $a_\mathrm{0,s}=0.005~\micron$,
$a_\mathrm{0,l}=0.1~\micron$, and $\sigma =0.75$, to roughly
cover the small and large grain size ranges.
We hereafter refer to this grain size distribution as
the lognormal model.
The normalizing constants are determined by
\begin{eqnarray}
\mu m_\mathrm{H}\mathcal{D}_i=
\int_0^\infty\frac{4}{3}\upi a^3sn_i(a)\,\mathrm{d}a,
\end{eqnarray}
where $\mu =1.4$ is the gas mass per hydrogen nucleus,
$m_\mathrm{H}$ is the hydrogen atom mass, and $s$ is
the material density of a dust grain. Therefore,
the evolutions of $\mathcal{D}_\mathrm{s}$ and
$\mathcal{D}_\mathrm{l}$ are reduced to those of
$C_\mathrm{s}$ and $C_\mathrm{l}$, respectively.

\subsection{Extinction curves}

The extinction at wavelength
$\lambda$ in units of magnitude ($A_\lambda$) normalized to
the column density of hydrogen nuclei ($N_\mathrm{H}$) is written
as
\begin{eqnarray}
\frac{A_\lambda}{N_\mathrm{H}}=2.5\log\mathrm{e}\sum_i\int_0^\infty
n_i(a)\upi a^2Q_\mathrm{ext}(a,\,\lambda ),
\end{eqnarray}
where $Q_\mathrm{ext}(a,\,\lambda )$ is the extinction efficiency
factor, which is evaluated by using the Mie theory
\citep{bohren83} and the same optical constants for
silicate and carbonaceous dust in
\citet{weingartner01}. For simplicity, the mass fractions of silicate
and carbonaceous
dust are assumed to be 0.54 and 0.46, respectively \citep{hirashita09}.
We adopt $s=3.5$ and 2.24 g cm$^{-3}$ for silicate and carbonaceous
dust, respectively \citep{weingartner01}.

{
In Appendix \ref{app:lognormal}, 
in order to examine how well the lognormal model is capable of
approximating the MRN model
($n_i(a)\propto a^{-3.5}$ with
$0.001~\micron \leq a\leq 0.25~\micron$),
which is known to reproduce the Milky Way extinction curve well,
we compare the extinction curves predicted by these two models
under the
same small-to-large grain abundance ratio. We confirm that the
lognormal model approximately reproduces the Milky Way extinction
curve, which means that the lognormal model provides a useful
simple approach to predict the extinction curve in the two-size
approximation.
}

\begin{figure}
\includegraphics[width=0.45\textwidth]{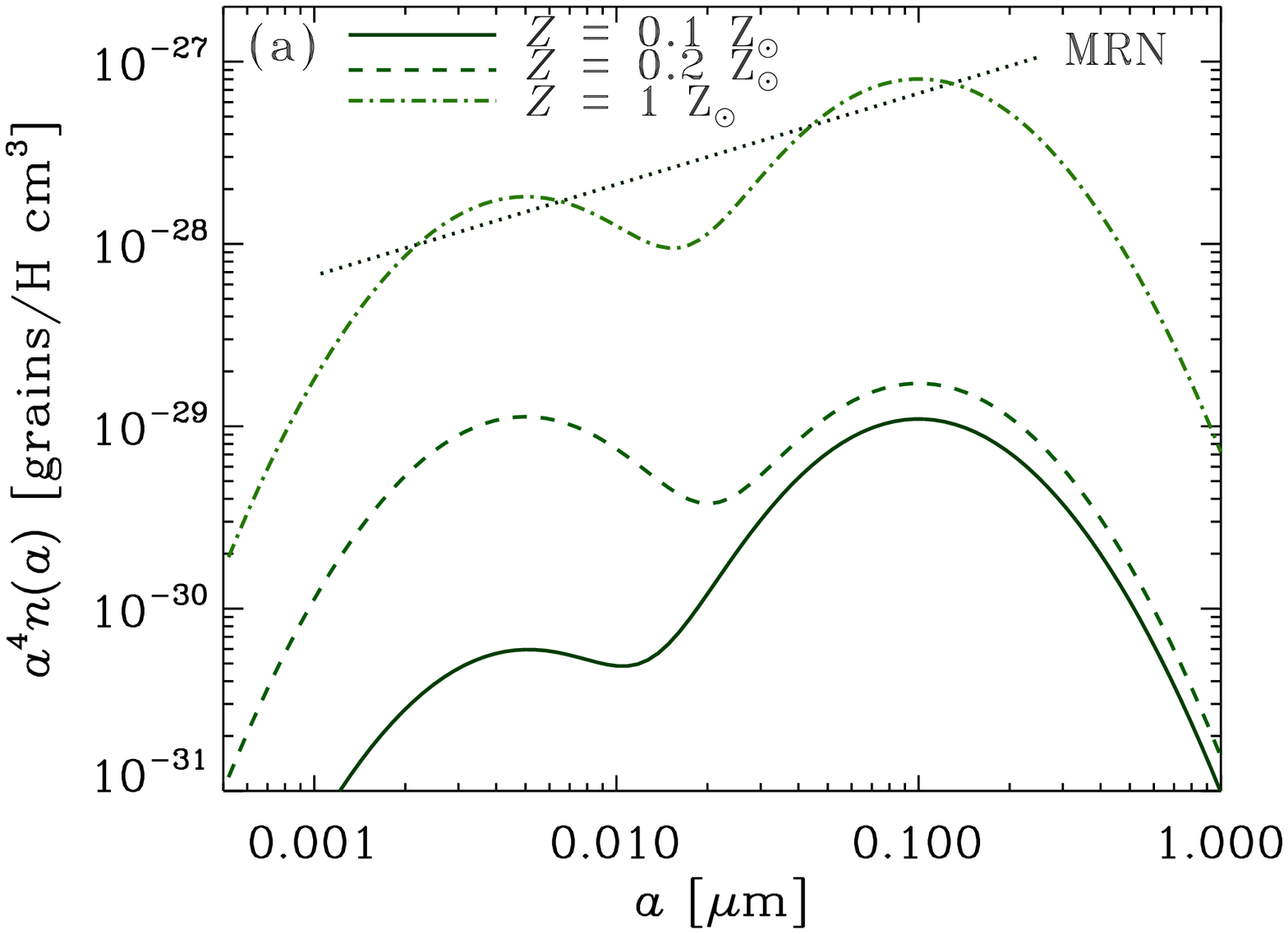}
\includegraphics[width=0.45\textwidth]{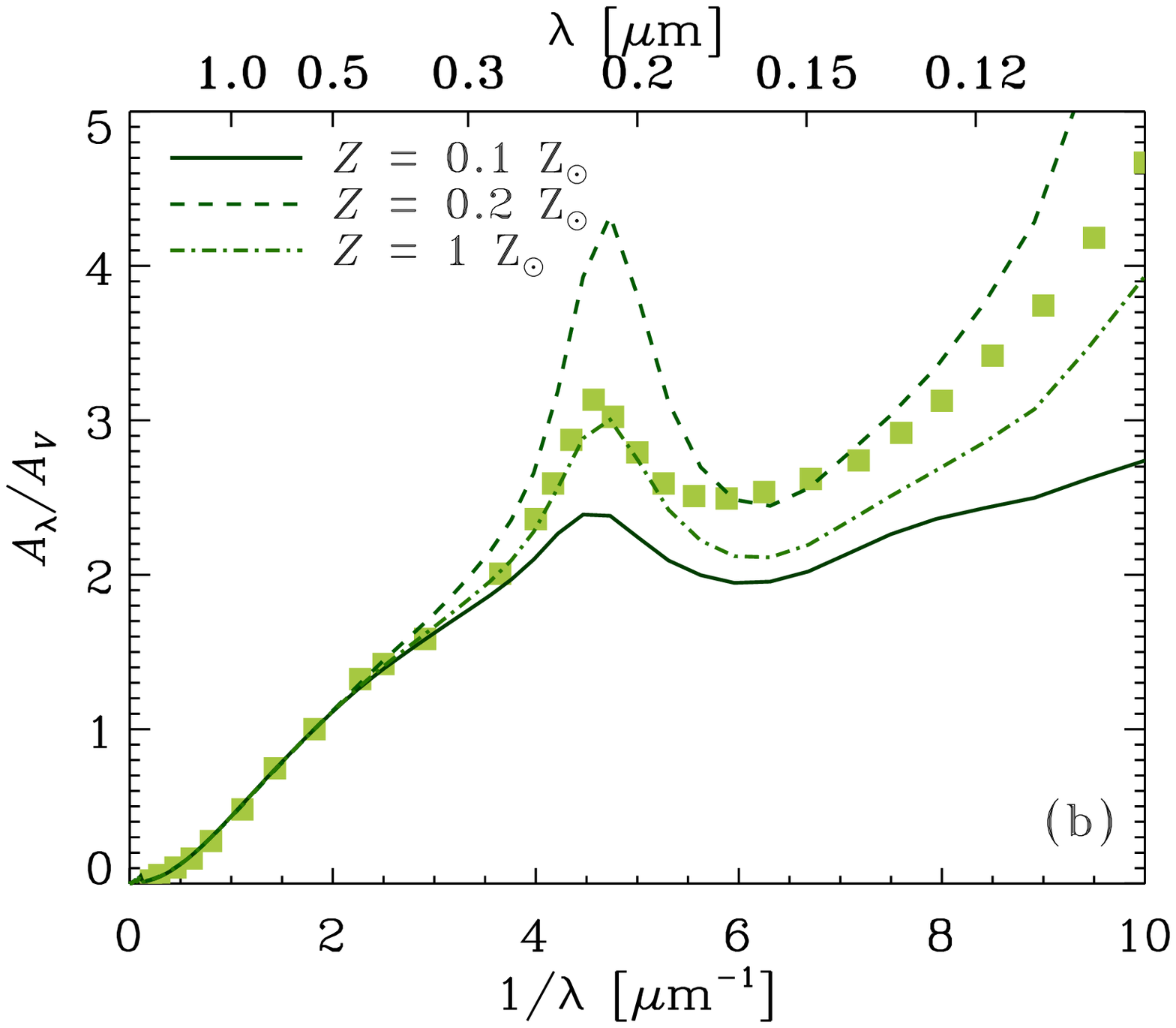}
\caption{
{(a) Grain size distributions in the lognormal
model for the fiducial model. The solid, dashed, and
dot-dashed lines show the size distributions at
$Z=0.1$, 0.2, and 1 Z$_{\sun}$, respectively.
The grain size distributions are presented by multiplying
$a^4$ to show the mass distribution per $\log a$.
The MRN grain size distribution is shown by the
dotted line as a reference.}
(b) Extinction curves calculated for the grain size distributions
in Panel (a). The extinction is normalized to the value at
the $V$ band (0.55 $\micron$). The solid, dashed, and dot-dashed lines show
the extinction curves at $Z=0.1$, 0.2, and 1 Z$_{\sun}$, respectively.
For reference, the filled squares show the observed mean extinction
curve of the Milky Way taken from \citet{pei92}.
\label{fig:ext_ev}}
\end{figure}

{
Now we examine the extinction curves
predicted by our two-size dust evolution models with the fiducial
parameter values.
In Fig.\ \ref{fig:ext_ev}, we show the grain size
distributions and
the extinction curves at
$Z=0.1$, 0.2, and 1 Z$_{\sun}$. At $Z\la 0.1$ Z$_{\sun}$,
the extinction curve is flat, reflecting the
grain size distribution dominated by the large grains.
At this metallicity, the dust content is
dominated by the stellar dust production of large grains.
At $Z\sim 0.2$ Z$_{\sun}$, the small-to-large grain abundance
ratio reaches its peak (Fig.\ \ref{fig:dust_metal}),
and indeed the lognormal model correctly show the enhanced
small-grain abundance. At this metallicity,}
the extinction curve becomes the steepest with the
most prominent carbon bump. Afterwards, coagulation
pushes the grains to large sizes, and accordingly the
extinction curve becomes flatter. The above evolutionary
features are in line with the evolution
of extinction curves calculated by A13's full treatment
of grain size distribution \citep{asano14}.

\begin{figure}
\includegraphics[width=0.45\textwidth]{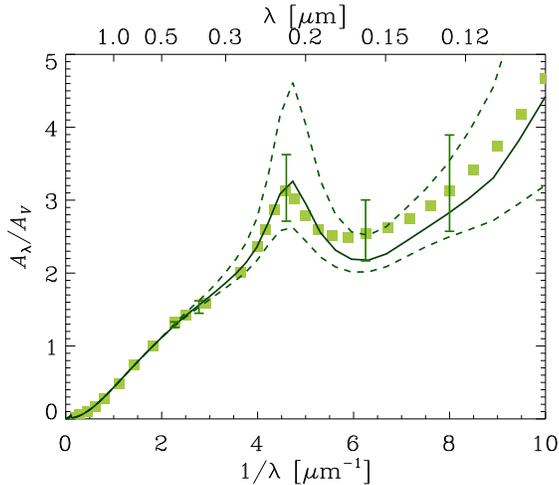}
\caption{
Extinction curves calculated with the lognormal model
for the mean (solid line), and the upper
and lower ranges expected for the dispersion (dotted lines)
in our Monte Carlo simulation.
The filled squares are the same
data as in Panel (a), and the vertical bars show the 1 $\sigma$
dispersion for the observed Milky Way extinction curves taken
from \citet{fitzpatrick07} \citep[see also][]{nozawa13}.
\label{fig:ext_disp}}
\end{figure}

We also examine the variation
at solar metallicity based on the dispersion of
$\log (\mathcal{D}_\mathrm{s}/\mathcal{D}_\mathrm{l})$
calculated by the Monte Carlo method in Section \ref{subsec:monte}
(see also Fig.\ \ref{fig:random}).
The mean and 1 $\sigma$ variation of
$\log (\mathcal{D}_\mathrm{s}/\mathcal{D}_\mathrm{l})$
is $-$0.52 and 0.40, respectively.
The extinction curves corresponding to the 1 $\sigma$ range
is shown in {Fig.~\ref{fig:ext_disp}}.
The variation of extinction curves is compared with the
variation in the Milky Way. The vertical bars in
Fig.\ \ref{fig:ext_ev}b show the 1 $\sigma$ range of
the Milky Way extinction curves in various lines of sight
compiled in \citet{fitzpatrick07} \citep[see also][]{nozawa13}.
The comparison between
the dispersion of the Milky Way extinction curves and
that predicted by our models assumes that the dispersion
is produced by the scatter of the parameters.
These two dispersions are roughly comparable except at the
carbon bump where our models predict too large a dispersion.
Probably the assumption that the silicate to graphite ratio
is constant should be revised in order to reproduce a reasonable
scatter of the carbon bump strength, or the optical properties of
carbonaceous dust may still need to be revised \citep{nozawa14}.

\subsection{Prospect for dust emission SED models}

Another important observational feature that reflects the
dust properties is the dust
emission SED at infrared and submillimetre wavelengths.
At long wavelengths
($\ga 100~\micron$), the dust responsible for the
emission can be treated as being in radiative equilibrium
with the ambient interstellar radiation field,
so that the SED shape depends mainly on the total dust
amount and the equilibrium dust temperature
\citep{desert90,li01}.
Therefore, the evolution of the grain size distribution
is not very important at those long wavelengths.

In contrast,
at wavelengths $\la 60~\micron$, the emission is governed
by small grains ($\la 0.01~\micron$), whose heat
capacities are so small that they are transiently
heated by discrete photon inputs
\citep{draine85}. Thus, the spectrum at $\la 60~\micron$ is
sensitive to the grain size distribution.
Moreover, some spectral features specific to
certain dust species  such as PAH features are prominent at
$\la 20~\micron$, which means that a detailed
spectral modeling is necessary for this wavelength range.
Therefore, the two-size models developed in this paper would be
too rough to predict the evolution of dust SED at such short
wavelengths.

Nevertheless, we could prepare spectral templates for
the small and large grains, and mix them according to the
small-to-large grain abundance ratio calculated by the two-size
models. If we separately calculate various grain species,
we could also sum up all the templates calculated for
various dust species. In this way, we will be able to
predict the evolution of dust emission SEDs without
consuming much computational time.

\section{Conclusion}\label{sec:conclusion}

We have examined a possibility of treating the evolution of
grain size distribution with two representing grain radii,
motivated by the fact that a full treatment of grain size
distribution is too heavy to be incorporated in cosmological
galaxy evolution models. In this `two-size approximation',
we consider
small (grain radius $a<0.03~\micron$) and large ($a>0.03~\micron$)
grains and treat the evolution of the mass fractions of
these two components through various processes of grain
formation and processing. We include dust supply from
stellar ejecta, destruction in SN shocks, dust growth
by accretion, grain growth by coagulation and grain disruption by
shattering. First we demonstrated that the following features of
the full treatment of grain size distribution by A13 are correctly
reproduced by the two-size approximation.
The dust enrichment starts with the supply of the large grains from
stars. The dominant dust production mechanism switches
to accretion at a metallicity level referred to as the critical metallicity
of accretion, since the abundance of the small grains formed by shattering
becomes large enough to rapidly increase the grain abundance by
accretion. Just after the system has reached this stage,
the small-to-large grain abundance ratio
reaches the maximum. After that, this ratio
converges to the value determined by the balance between
shattering and
coagulation, and the dust-to-metal ratio is determined by the
balance between accretion and shock destruction.

With a Monte Carlo simulation, we showed that the simplicity
of our model has an advantage in predicting statistical properties.
We also demonstrated that our two-size approximation
is also useful in predicting observational dust
properties such as extinction curves. In our future work,
we will incorporate the formulation developed in this paper
into cosmological galaxy formation and evolution models.

\section*{Acknowledgments}

{We thank R. Schneider for careful reading and helpful comments.}
We are grateful to A. R\'{e}my-Ruyer for providing us with
the nearby galaxy data of dust-to-gas ratio and metallicity, and
to Y.-H. Chu, {K.-C. Hou}, and the interstellar and circumstellar group
in our institute for useful comments.
We thank the support from the Ministry of Science and Technology
(MoST) grant 102-2119-M-001-006-MY3.

\appendix

\section{Test for the lognormal model}\label{app:lognormal}

{
We examine how well the Milky Way extinction curve is
reproduced by the log-normal model of grain size distribution
(equation \ref{eq:lognormal}), comparing it with the
\citet{mathis77} (MRN) grain size distribution
($\propto a^{-3.5}$ in the grain radius range of
0.001--0.25 $\micron$), which is already known to
reproduce the Milky Way extinction curve well.
The mass ratio of silicate to carbonaceous dust is assumed to
be the same as the value used in the text (0.54 : 0.46).
We adopt the same small-to-large grain abundance ratio as the
MRN model (0.30 : 0.70; see section \ref{subsec:evol});
that is, we do not use the output of our dust
enrichment model for the purpose of simply comparing with
the MRN model.}

\begin{figure}
\includegraphics[width=0.45\textwidth]{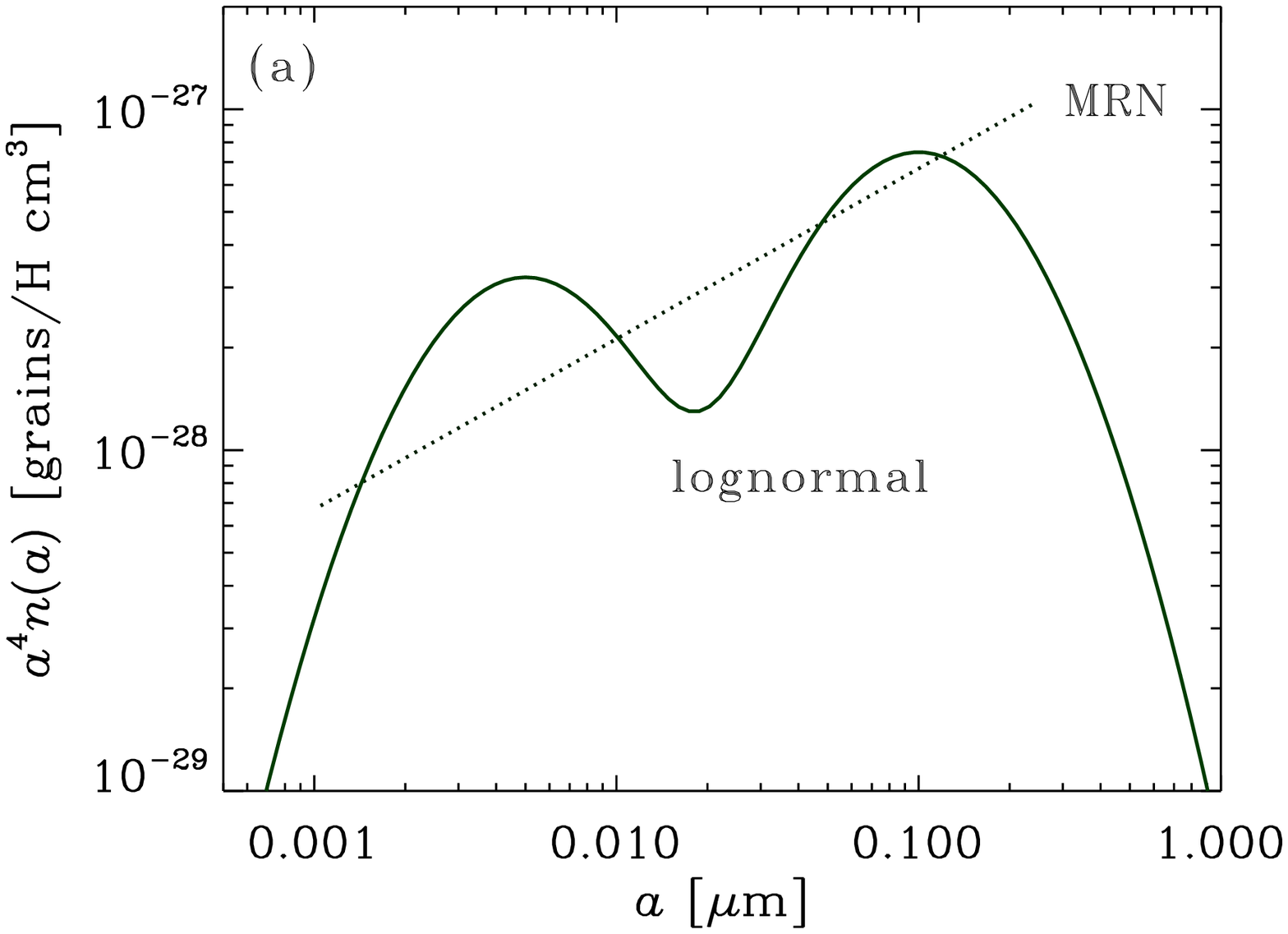}
\includegraphics[width=0.45\textwidth]{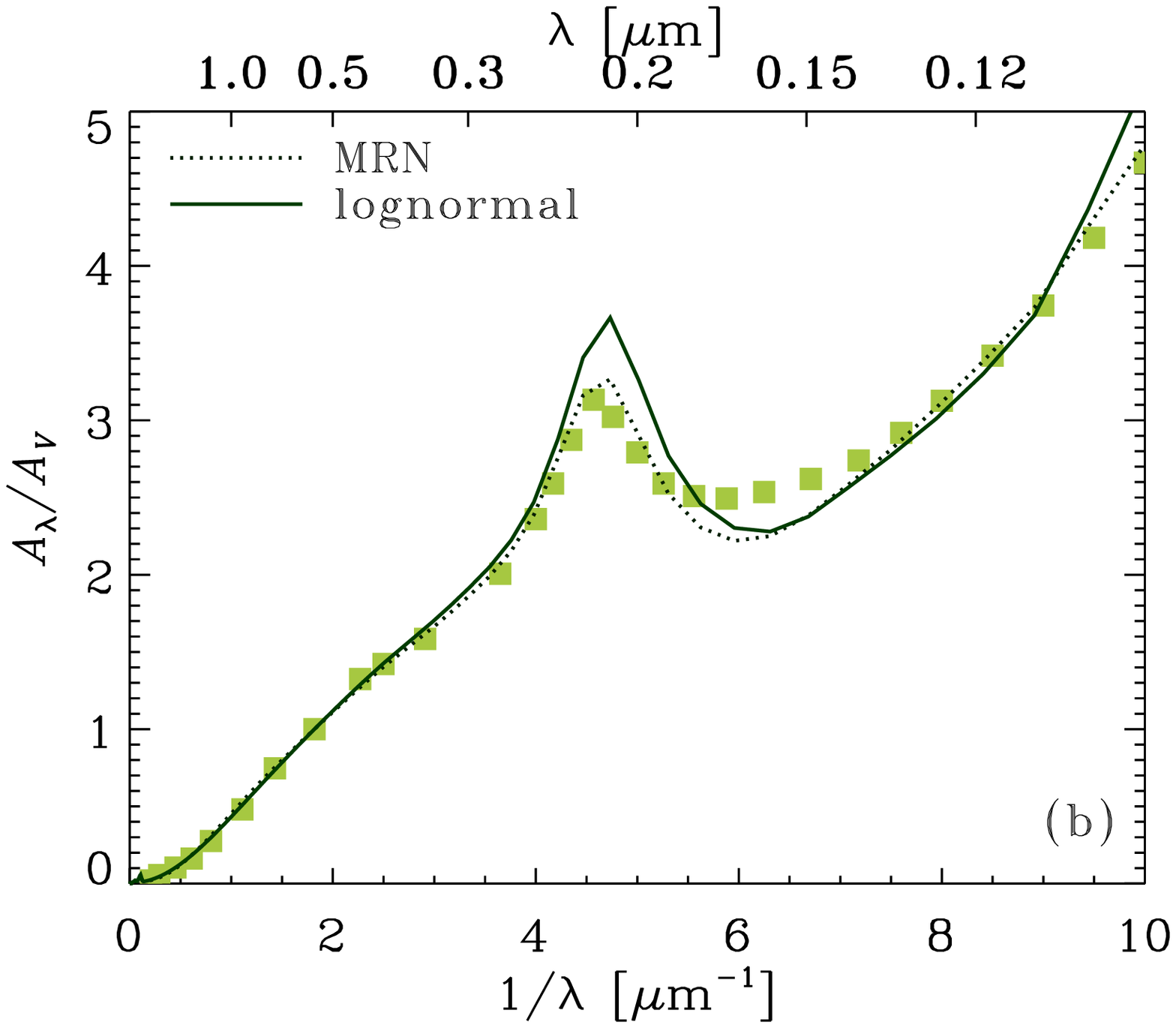}
\caption{(a) Comparison between the two grain size distribution
models (MRN and lognormal models)
{with the same small-to-large grain abundance ratio}.
The grain size distributions
are presented by multiplying $a^4$ to show the mass distribution
in each logarithmic bin of the grain radius $a$.
(b) Extinction curves calculated with the lognormal and MRN
models (solid and dotted lines, respectively). The extinction
is normalized to the value at the $V$ band (0.55 $\micron$).
The filled squares are the observed mean extinction curve
of the Milky Way taken from \citet{pei92}.
\label{fig:ext}}
\end{figure}

{
In Fig.\ \ref{fig:ext}, we compare the grain size distributions and
the extinction curves.
Although the lognormal model deviates from
the observed
mean extinction curve of the Milky Way more than the MRN model,
the difference between these two models is significant only
at the bump around $\sim 0.22~\micron$ contributed from small
graphite grains and
toward far-ultraviolet wavelengths, and the deviations are
within 20 per cent. We expect that, in calculating
extinction curves in extragalactic objects, the uncertainties
arising from assumed grain properties are rather larger.
Therefore, we conclude that the simple lognormal model
provides a good approximate method of calculating extinction
curves based on the two-size approximation.
}

\bsp

\label{lastpage}

\end{document}